\documentclass[a4paper,verb|alph-refs| ]{aejstyles}

\journal{aej}

\usepackage{graphicx}
\usepackage{siunitx}
\usepackage{amsmath}
\usepackage{amsfonts}
\usepackage{siunitx}
\usepackage{pdfpages}
\usepackage{appendix}

\title{Status of Astronomy Education in India: A Baseline Survey}

\author[1, 2, \authfn{1}]{Moupiya Maji}
\author[1, 2]{Surhud More}
\author[2, 3]{Aniket Sule}
\author[4]{Vishaak Balasubramanya}
\author[1,2]{Ankit Bhandari}
\author[5]{Hum Chand}
\author[1,2]{Kshitij Chavan}
\author[6,7]{Avik Dasgupta}
\author[8]{Anindya De}
\author[9,10]{Jayant Gangopadhyay}
\author[11]{Mamta Gulati}
\author[12]{Priya Hasan}
\author[13,14]{Syed Ishtiyaq}
\author[12]{Meraj Madani}
\author[15]{Kuntal Misra}
\author[4]{Amoghavarsha N}
\author[16]{Divya Oberoi}
\author[17]{Subhendu Pattnaik}
\author[1]{Mayuri Patwardhan}
\author[4]{Niruj Mohan Ramanujam}
\author[3]{Pritesh Ranadive}
\author[18]{Disha Sawant}
\author[5]{Paryag Sharma}
\author[11]{Twinkle Sharma}
\author[3]{Sai Shetye}
\author[2,3]{Akshat Singhal}
\author[19]{Ajit M. Srivastava}
\author[5]{Madhu Sudan}
\author[20]{Mumtaz Syed}
\author[4]{Pulamathi Vikranth}
\author[15]{Virendra Yadav}

\affil[1]{Inter University Center for Astronomy and Astrophysics, Pune}
\affil[2]{IAU-OAE Center India}
\affil[3]{Homi Bhabha Center for Science Education, Mumbai}
\affil[4]{Indian Institute of Astrophysics, Bengaluru}
\affil[5]{Central University of Himachal Pradesh, Dharamshala}
\affil[6]{Vikram A Sarabhai Community Science Centre, Ahmedabad}
\affil[7]{Centre for Cosmology and Science Popularization, Gurugram}
\affil[8]{Hindu School, Kolkata}
\affil[9]{Regional Science Centre and Planetarium, Calicut}
\affil[10]{Mahakushal University }
\affil[11]{Thapar Institute of Engineering and Technology, Patiala}
\affil[12]{Maulana Azad National Urdu University,  Hyderabad}
\affil[13]{Department of school Education, Jammu \& Kashmir}
\affil[14]{National Institute of Technology, Srinagar}
\affil[15]{Aryabhatta Research Institute of Observational Sciences, Nainital}
\affil[16]{National Centre for Radio Astrophysics, Pune}
\affil[17]{Pathani Samanta Planetarium, Science and Technology Department, Govt. of Odisha}
\affil[18]{Pune Knowledge Cluster, Pune}
\affil[19]{Institute of Physics, Bhubaneswar}
\affil[20]{The Sky Explorers, Mumbai}

\authnote{\authfn{1}moupiya.maji@iucaa.in}
\papercat{Research Article}

\runningauthor{Maji et al.}

\jvolume{00}
\jnumber{0}
\jyear{0000}

\begin{document}

\begin{frontmatter}
\maketitle
\begin{abstract}
%The Abstract (250 words maximum) should be structured to include the following details: \textbf{Background}, the context and purpose of the study; \textbf{Results}, the main findings; \textbf{Conclusions}, brief summary and potential implications. Please minimize the use of abbreviations and do not cite references in the abstract.

We present the results of a nation-wide baseline survey, conducted by us, for the status of Astronomy education among secondary school students in India. The survey was administered in 10 different languages to over 2000 students from diverse backgrounds, and it explored multiple facets of their perspectives on astronomy. The topics included students' views on the incorporation of astronomy in curricula, their grasp of fundamental astronomical concepts, access to educational resources, cultural connections to astronomy, and their levels of interest and aspirations in the subject. We find notable deficiencies in students' knowledge of basic astronomical principles, with only a minority demonstrating proficiency in key areas such as celestial sizes, distances, and lunar phases. Furthermore, access to resources such as telescopes and planetariums remain limited across the country. Despite these challenges, a significant majority of students expressed a keen interest in astronomy. We further analyze the data along socioeconomic and gender lines. Particularly striking were the socioeconomic disparities, with students from resource-poor backgrounds often having lower levels of access and proficiency. Some differences were observed between genders, although not very pronounced. The insights gleaned from this study hold valuable implications for the development of a more robust astronomy curriculum and the design of effective teacher training programs in the future.

\end{abstract}

%\vspace{10cm}


\end{frontmatter}

\section{Introduction}
\label{sec:intro}
Astronomy is consistently regarded as one of the most fascinating subjects by people of all ages, and this is especially true for children of school age. Any interactions with groups of school students either in formal (such as school classrooms), informal (such as public talks, science outreach programs) and non-formal(such as campus visits, group chats etc.)
settings reveal that a large majority of students are extremely curious about astronomical topics. Even though in popular culture, astronomy is regarded as only the realm of the most extreme objects (e.g. black holes) or mind-bending theories (e.g. parallel universes), the basics of astronomy are necessarily rooted in physics, chemistry, and mathematics. Thus astronomy can be an extremely useful subject to teach at the school level \citep{Percy2005}, not only because of its unique appeal among young people but also because it can be used to teach science subjects better, demonstrate the interdisciplinary aspects of science, attract students towards STEM education and above all provide students a broader perspective. In this sense, it is often described as a ``gateway science''. 

% overview of the situation worldwide - is astronomy included in school? --------------
Despite its profound potential, astronomy remains significantly underutilized in school curricula worldwide, presenting two interconnected challenges. Firstly, the extent of use of astronomy within educational frameworks varies drastically among nations, with a few including it as a standalone subject or integrating within science disciplines, while others allocating minimal attention to astronomical concepts. \cite{Salimpour2021} reviewed 52 curricula (grades 1 - 12) from 37 countries including 35 OECD (Organisation for Economic and Cooperative Development) member countries (most of them are developed countries from the Americas, Europe, and Pacific), China and South Africa, and found that there is some astronomy-related content in all of these curricula, and that they are most commonly covered in grade 6 and 17$\%$ of curricula offered astronomy as an optional subject. However, they find that only 27$\%$ curricula explicitly mention astronomy content in all grades (1 - 12) and most of the content across all curricula is descriptive rather than conceptual. In the USA, a standalone high school astronomy course can be found in about $10 - 12\%$ of schools \citep{Krumenaker2009}. For the rest, it is mostly a small part of courses on either earth sciences or physics.   

% how much of it students really learn? ------------------
Secondly, even for the astronomy topics present in the curricula, the depth of student comprehension is often limited. Research examining the student's understanding of astronomy can be categorized into two distinct methodologies: surveys or general tests, which encompass inquiries spanning various topics to provide a comprehensive overview, and concept inventories (CIs), which employ targeted questions focusing on specific concepts to elucidate a deeper understanding of that particular subject matter. In the literature, most studies have focused on either elementary school levels \citep{Vosniadou1994, Vosniadou1992, Nussbaum1976} or undergraduate levels \citep{Slater2014, ADT2002}. There has been relatively less focus on middle and high-school-level topics. A review of the literature \citep{bigideas2010} shows that most middle and high school students find the concepts related to lunar phases, seasons, the earth-sun-moon system, and gravity difficult to understand. 

% more literature surveys from around the world.
In particular, \citep{Sadler2009} developed and used the ASSCI (Astronomy and Space Science Concept Inventory) on 787 middle and 249 high school students in USA, and found that students have many misconceptions about astronomical objects (Sun, solar system) and phenomena (moon phases, seasons, etc.). They also found that after a year of study with some emphasis on astronomy, middle school students achieved only small gains (small difference between pre and post-course tests) whereas, after an astronomy course, high school students reported significant gains. \citep{Slater2018Australia} conducted a survey with 546 students in Australia using a modified version of the Astronomy Diagnostic Test (ADT, \citet{ADT2002}) and found that students held a multitude of misconceptions about the commonly taught astronomy topics. \cite{Trumper2001} surveyed 378 high school students in Israel about similar astronomy topics and found that the overall correct response rate was $43.6\%$. 
In addition to the aforementioned topics, research indicates that students encounter considerable challenges in comprehending astronomical scales. For instance, a study conducted by \citep{Rajpaul2018} involving 922 middle and high school students in Norway revealed that a majority of students struggle to understand the sizes and distances of celestial bodies, even after instruction. Pedagogical research into intervention strategies suggests that to successfully impart the correct scientific understanding of a topic, educators must explicitly address students' misconceptions or mental models during instruction \citep[and references within]{bigideas2010}.

% Position of Astronomy in the Indian education system
India has a rich cultural heritage of practicing astronomy. One of the earliest texts, the Vedanga Jyotisha, dating back to around 1400 BCE, demonstrates the significance of celestial observations for calendrical and religious purposes. Throughout the centuries, Indian astronomers made significant advancements in observational astronomy, constructing sophisticated instruments (e.g. Astrolabe, Gnomon) and later, observatories (e.g. Jantar Mantar). Almost all the Indian festivals are intricately linked to particular lunar phases. Despite this, the present Indian education system provides very limited opportunities for learning basic astronomy. In schools, astronomy does not appear in the curriculum as a separate subject. Instead, it is sporadically introduced across various disciplines, primarily integrated within subjects like environmental science and geography, in middle school. It is worth noting that India's educational landscape is diverse, with multiple school boards administering their respective curricula across the nation's 28 states, alongside a few central boards. These states also encompass a multitude of languages, contributing to significant linguistic variety within the school systems. Despite this diversity, the majority of these boards broadly adhere to the syllabus outlined by the National Council of Educational Research and Training (NCERT\footnote{https://ncert.nic.in/}). In the NCERT curriculum, the astronomy topics are introduced in grades spanning from 5 to 8 and include concepts about the motions of the Earth, the solar system, the night sky (stars and constellations), seasons, phases of the Moon, eclipses, and tides. As students progress to grades 9 through 12, astronomy topics make occasional appearances in physics books, e.g. gravitation, Kepler's laws of planetary motion, optics, and satellites.

% Situation of AER in India
Astronomy Education Research (AER) is still in a nascent stage in India. There have been few studies on intervention methods for students, for example, \cite{Padalkar2011} demonstrated the positive role of gestures and actions in learning the sun-earth-moon system, and \cite{Subramaniam2009} examined the role of visualization in understanding moon phases. As a part of the International Astronomy Union's (IAU) Office of Astronomy for Education (OAE) Center India\footnote{https://astro4edu.iucaa.in/}, we strive to improve astronomy education in Indian schools.  The first step towards this is assessing the current situation, i.e. the state of knowledge and attitude of students towards astronomy at present. However, there have been no studies to capture the broader picture of astronomy education in India. Therefore, in this project, we have conducted a nationwide survey about astronomy education with school students. This is the first large-scale survey of its kind in India.

% why this survey is important
% What exact research questions do we focus on?
In this work, we focus on the following research questions:
\begin{itemize}
    \item To what extent does the current astronomy content in Indian schools effectively reach students?
    \item  What are the expectations and aspirations of students regarding astronomy?
    \item Can students effectively connect their acquired knowledge of astronomy with their daily lives or their cultural heritage?
    \item Are there any differences in students' responses based on their gender and/or socioeconomic status?
\end{itemize}

This paper is organized as follows: in $\S$~\ref{sec:methods}, we describe the methodology of our survey, we discuss the findings of the survey in $\S$~\ref{sec:results}, and finally we discuss the implications and limitations of our work and present our inferences in $\S$~\ref{sec:discussion}.

\section{Methodology} %---------
\label{sec:methods}
We began this study by designing the survey questionnaire, followed by pilot testing and subsequent revisions to enhance its efficacy. Then the finalized survey questionnaire was administered nationwide with the help of a team of collaborators. After data collection, responses were meticulously encoded and digitized for systematic analysis. Below, we describe each of the steps in detail.

\subsection{Survey questionnaire}
This survey is intended to be a general-purpose questionnaire to get a broad sense of students' knowledge and attitudes toward astronomy. The initial versions of the survey questions were inspired by many discussions with experienced teachers, students, outreach professionals, and education researchers. We also consulted other AER surveys and tests, e.g. Astronomy Diagnostic Test \citep{ADT2002} and the Astronomy and Space Science concept inventory \citep{Sadler2009}. During the development phase, particular attention was paid to ensure that the questions reflected the Indian context appropriately. 

Our survey instrument consists of 16 questions (\S~\ref{sec:appendix}). It is important to note that all of them are designed to be short answer type questions, where students are encouraged to experss themselves freely without any constraints or guidance. This is in contrast to most diagnostic tests and CIs, which often tend to be multiple-choice questions (MCQs). The short-answer type questions can elicit more variety and in-depth answers from respondents. They can also give us a more realistic picture of students' knowledge as they are not required to choose from given choices. This is particularly important, as no prior similar studies on student thinking have been conducted in India and hence pre-selecting and presenting only few choices to the respondents would not be optimal. On the other hand, analyzing the free-hand answers of a large number of students is far more time-consuming than a MCQ test. Transcribing the student's answers to a coded form was the most time-consuming part of this project for our team.

The questions in our survey can be broadly divided into five themes: astronomy in the curriculum, general astronomy knowledge, cultural connection, exposure to astronomy, and their interest in astronomy. In the first three questions, we ask the students about their favorite subject, if they like astronomy and the most liked/least liked/hardest astronomy topics in their school curriculum. The next three questions probe their general astronomy knowledge about distance scales and mass scales together with moon phases. In the next two questions, we explore their cultural connection with astronomy, e.g. if they can name festivals that are linked to new moon / full moon days (most Indian festivals follow the lunar calendar) in addition to probing their belief in astrology.  Then we inquire about their exposure to astronomy with questions about access to the night sky, telescopes, planetarium, and other astronomy-related resources. The last few questions are about their further interest, i.e. interests in other astronomy topics, related courses in higher study, and astronomy as a career.

\subsection{Pilot study}
In the pilot phase, the survey was administered to students studying in grade 9 in India\footnote{Grade 9 in India is equivalent to the first year or freshman year of a 4-year high school in the US system}. Typically, students within this cohort are approximately 14 to 15 years old. We chose this grade for the survey because, in the majority of Indian school boards, all major astronomy content is covered by the end of class VIII, as discussed in $\S$~\ref{sec:intro}. Therefore, we argue that at the beginning of 9th grade students should have both knowledge and retention of astronomy concepts.

In preparation for the nationwide implementation of our survey, a pilot study was conducted in Pune city in English. For this purpose, we selected two urban schools where the survey was administered to 268 grade IX students. Both schools were English medium and followed the state board curriculum. 

We used the data from the pilot survey to revise the questionnaire to enhance the survey instrument's efficacy. We placed particular emphasis on avoiding scientific jargon as much as possible and where unavoidable, we either defined the terms (e.g. astronomy) in the survey or provided a vernacular translation in parenthesis (e.g. full/new moon, horoscope). Besides, a deliberate effort was made to structure questions in a manner that focused on singular concepts, mitigating potential confusion or ambiguity in the responses. For example, question 16 asked students about the course of study required to become an astronomer to gauge their awareness of this career path. The pilot survey responses found that students gave very general and vague answers (e.g. study science/astronomy) and probably misunderstood the question. So based on this feedback, we made the question more targeted to ask the path in definite steps, i.e. the course of study required after 10th class, 12th class, college, and beyond. We also thoroughly checked for readability and accessibility of the questions. After a series of iterative modifications and refinements, a finalized version of the survey was obtained which is presented in its entirety in Appendix 1.

\subsection{Encoding Scheme}
\label{sec:coding}
Since all of the survey questions were free-form, i.e. short answer type, we needed a coding scheme to systematically digitize the data for analysis. We used the pilot survey responses to develop such an encoding scheme. The scheme also helped us to unify the responses obtained in multiple languages.

For the open-ended questions, we developed categories based on the pilot responses and later condensed them into keywords which were used to digitize the data in the full survey. The responses to the close-ended questions were categorized as yes, no, maybe or correct, partially correct, and wrong, as applicable. The answers to knowledge-based questions (questions 4,5 and 6) were recorded as is. For all questions, the answers left blank were tagged as NA. When analyzing each question, we discarded the NA responses unless otherwise specified. As different numbers of pupils left different answers blank, the total number of valid responses can vary for the different questions.

The exact details of the encoding scheme will be mentioned for every question while discussing the results of the survey.

Our collaborators were given training to administer the survey and then encode the survey responses. We ensured that all collaborators contacted the schools and got the necessary permissions in advance. The text for the consent letters for the school principals and teachers were shared with the whole team. All the collaborators made sure to explain the consent process to students in class before the survey, in the language of instruction of the school. The encoding process was written as a document which was discussed and shared with the collaborators.

\subsection{Translations of the survey}

India, a nation characterized by extensive linguistic diversity with twenty-two major languages, presents a unique challenge in the context of survey research. Although a significant percentage of schools conduct lessons in English ($~ 26\%$ students are enrolled in English medium school, UDISE report 2019-2020), most schools across the country use vernacular medium. In addition, accessibility to English medium schools can also bias the cohort population in terms of socio-economic background. Therefore to avoid such biases and to optimize accessibility for the student demographic, it was imperative to conduct the survey in the linguistic medium employed by the individual schools participating in our study. To achieve this, our team of collaborators took a major initiative to translate the survey into the language of their respective states. Thus, the survey has been meticulously translated into nine languages, namely Marathi, Hindi, Bengali, Kannada, Malayalam, Telugu, Punjabi, Gujarati, and Odia. This multilingual approach ensures that all students can engage with the survey in a language familiar to them, thereby mitigating potential language-related barriers and fostering more equitable and inclusive participation across the nation.

It is important to note that, unlike English, in most Indian languages verbs have separate masculine and feminine forms. Thus, special attention was paid to ensure that the translations remain gender neutral. e.g. every sentence addressed to the responder included both forms of the corresponding verbs.

% anything about the Romanian translation? Last I checked, she had done the translation.

\subsection{Survey sample} %---------

\begin{figure}
    \centering
    \includegraphics[width=0.45\textwidth]{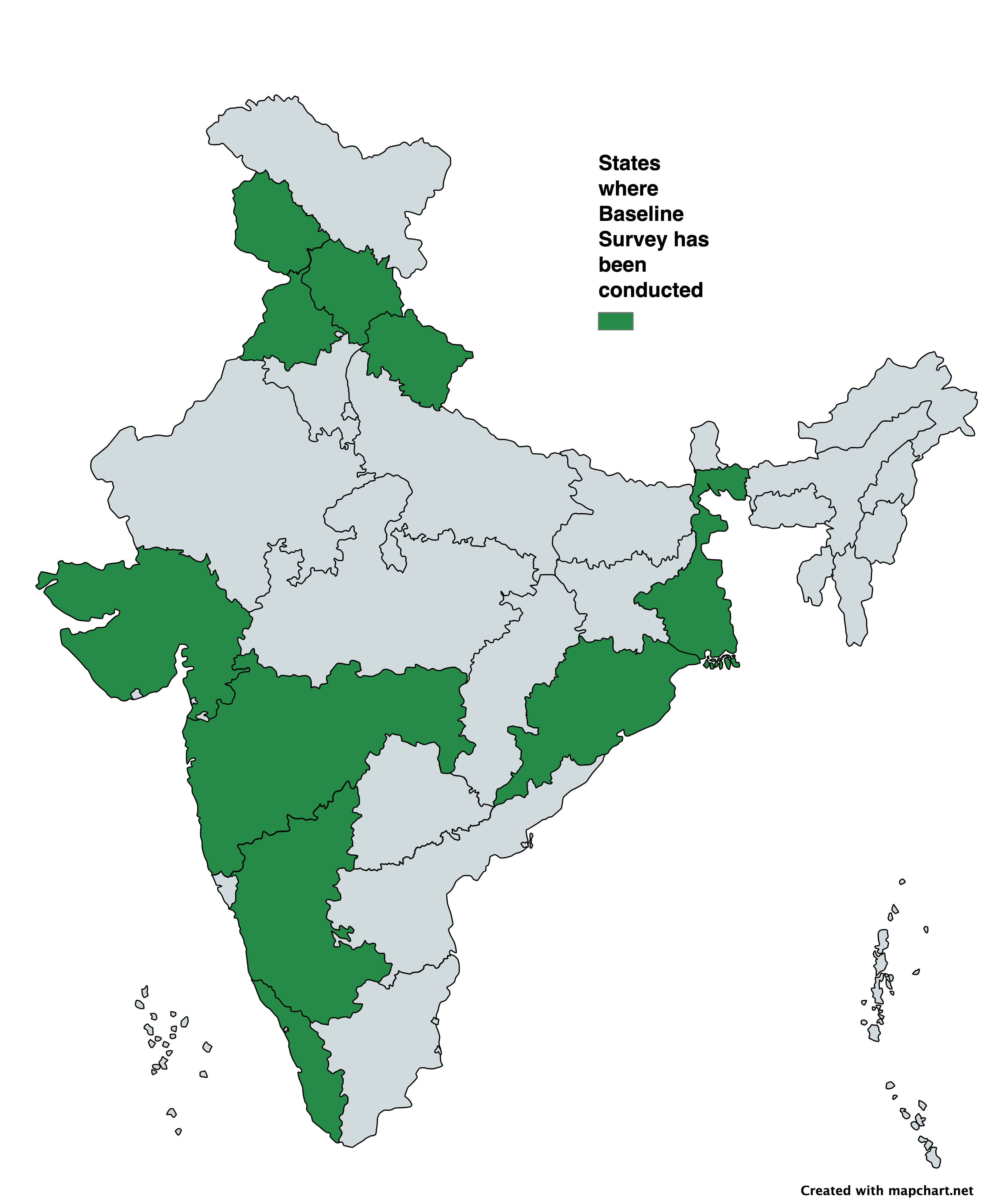}
    \caption{Map of India showing the states (in green color) where the Baseline Survey has been conducted.}
    \label{fig:map}
\end{figure}

In a large and diverse country such as India, we need large-scale surveys to assess the status of astronomy education within the country. Consequently, our survey endeavors to encapsulate a spectrum of settings, spanning urban, semi-urban, and rural regions, alongside schools affiliated with diverse curriculum boards and instructional mediums. We have conducted this survey in ten different states across the country (Figure~\ref{fig:map}), namely Maharastra, Gujarat, Kerala, Karnataka, West Bengal, Uttarakhand, Punjab, Himachal Pradesh, Kashmir and Odisha\footnote{The survey was also conducted in Telengana, but we later realized that due to some miscommunication, in most of these schools the survey was conducted post an introductory astronomy session where some of the content covered in the questionnaire may have been discussed. This raised significant concerns about biases, therefore we have excluded this data from our analysis.}
%Some of our intended collaborators conducted the survey in Telangana (more than 100 students in 2 schools) as well; however, at the analysis stage, the patterns in the data from this state raised significant concerns about data validity and integrity. Therefore, we excluded this data from our analysis}.
The details of the survey sample are described in Table 1. 

\begin{table*}
    \centering
\caption{Details of the survey sample. The rows show the states in which the survey was administered, the aggregate count of schools that participated in each state, the distribution of schools into categories of resource-rich and resource-poor, the total number of student respondents, and the demographic breakdown of male and female students. The last row shows the total numbers nationwide.}
\label{tab:my_label}
    \begin{tabular}{|c|c|c|c|c|c|c|} \hline 
         State                & Total schools & Resource Rich &  Resource Poor & Total Students & Male & Female\\ \hline 
         Maharashtra          & 14 & 3 & 11 & 939 & 399 & 466 \\ \hline
         Gujarat              & 3 & 1 & 2 & 143 & 74 & 69 \\ \hline
         Himachal Pradesh     & 2 & 1 & 1 & 115 & 73 &42 \\ \hline
         Jammu \& Kashmir     & 2 & 1 & 1 & 163 & 50 & 113 \\ \hline
         Karnataka            & 2 & 1 & 1 & 210 & 95 & 104\\ \hline
         Kerala               & 2 & 1 & 1 & 89 & 24 & 65 \\ \hline
         Odisha               & 2 & 1 & 1 & 107 & 46 & 45 \\ \hline
         Punjab               & 1 & 1 & 0 & 89 & 54 & 35 \\ \hline
         Uttarakhand          & 4 & 1 & 3 & 97 & 25 & 64 \\ \hline
         West Bengal          & 2 & 1 & 1 & 86 & 37 & 49 \\ \hline \hline
         Nationwide           & 34 & 12 & 22 & 2038 & 871 & 1014 \\ \hline 
         
    \end{tabular}
\end{table*}

In total, our survey encompasses a cohort of 2038 students across 34 schools. As the majority of our research team is based in Maharashtra, we have been afforded greater access to a diverse array of schools here. Consequently, we have surveyed 14 schools in this state. Our collaborative partners in the remaining 10 states predominantly conducted surveys in two schools each, while some states conducted assessments in three or four schools (with one state featuring one school).

\begin{figure*}
    \centering
    \includegraphics[width=0.9\textwidth]{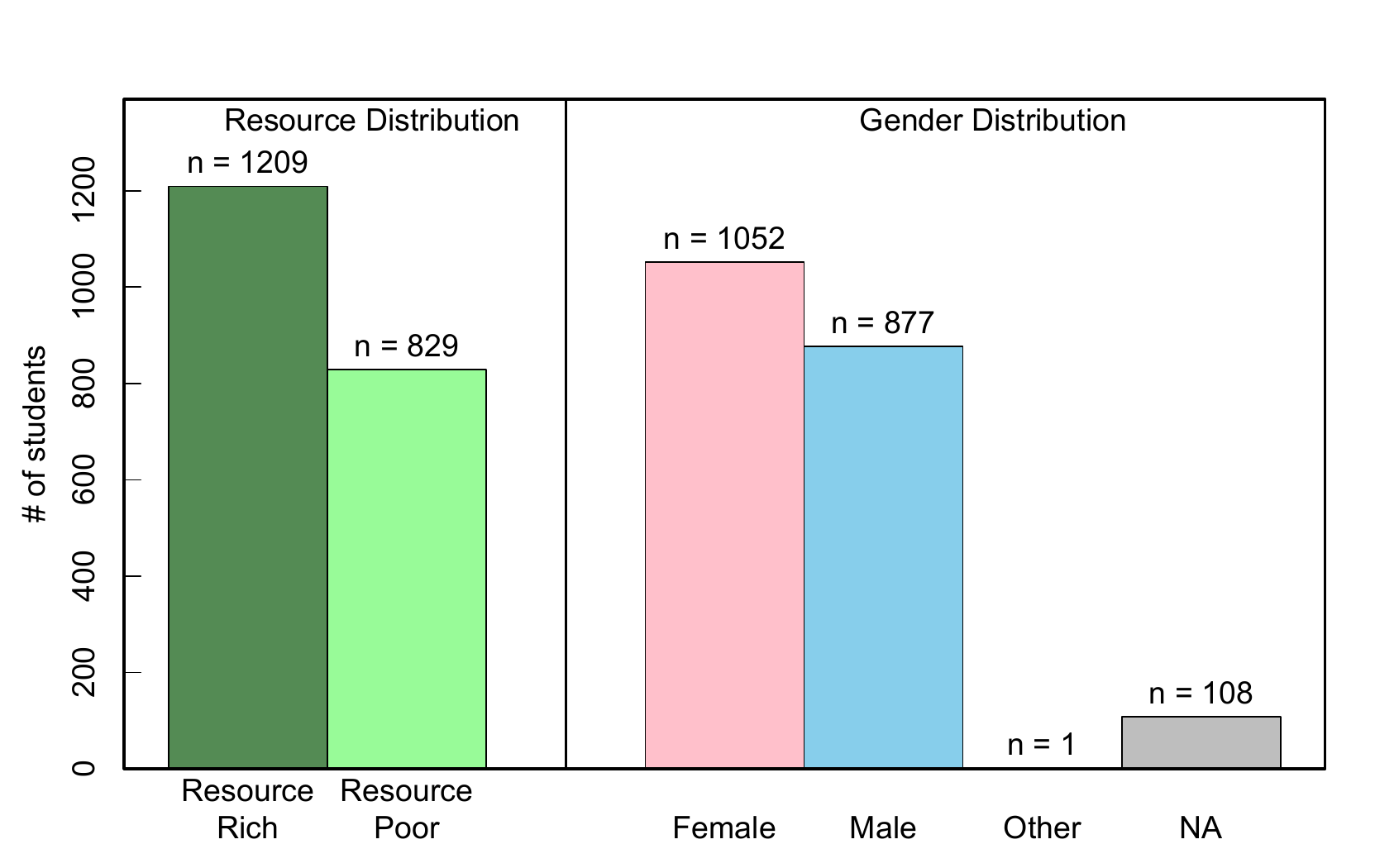}
    \caption{Histograms depicting the resource distribution (left) and the gender distribution (right) of the respondents in the survey.}
    \label{fig:gender_hist}
\end{figure*}

In every state, we have ensured to visit at least one urban school and one rural school. During the field survey, we realized that there is a large disparity among resources available to schools. Although some schools are resource-wise affluent, many have very limited resources irrespective of their urban or rural setting. Here resources may mean access to library, laboratory facilities, hands-on activity kits, science models, computers, etc. Even the availability of teachers can be inadequate in many schools. So we have divided the schools into two categories for a more meaningful discussion: \textbf{resource-rich} and \textbf{resource-poor}. We have not articulated any particular guidelines to denote a school to be resource-rich or poor, rather we have asked each of our collaborators who visited the schools to categorize them based on their first-hand experience. This characterization was done prior to any analysis of actual data to avoid any biases. Among our respondents, $59\%$, or 1209 students attend resource-poor schools, and $41\%$, or 829 students attend resource-rich schools (Figure~\ref{fig:gender_hist}).

We have also collected some basic demographic data from the survey respondents e.g. their gender (Figure~\ref{fig:gender_hist}). We find that for 108 or $5\%$ respondents the gender data is not available (i.e. students left the field blank). Among the rest, there is a slightly higher percentage of female students ($1052$ students or $55\%$) compared to the male student population (877 students or $45\%$). One student has mentioned their gender as Other. Since the other category has only one response, in our subsequent analysis of gender-related differences, we have used data from only male and female students.

\subsection{Statistical analysis}
Beyond an overall analysis of the data, in this paper, we will further discuss the results of our survey along two different demographic lines: gender, i.e. male vs female, and socioeconomic status, i.e. students from resource-rich and resource-poor schools. To compare results among these different groups, we will mostly use the chi-square test.

The chi-squared test is a widely used statistical method employed to examine the relationship between two categorical variables. Unlike tests designed for continuous data, such as t-tests or ANOVA, the chi-squared test is particularly suited for analyzing categorical variables that encompass two or more distinct categories. In the context of our study, where the focus often lies in determining the correctness of responses to specific questions, our data predominantly falls under the categorical domain. When comparing two sets of variables, such as gender vs performance, each with binary categories, male/female and correct/incorrect responses, the chi-squared test enables the assessment of whether there is a significant relationship between them.

The null hypothesis in these tests is that there is no association between the two categorical variables. The chi-square test essentially checks if the observed frequency of the categories is significantly different from what would be expected from the null hypothesis. By calculating the chi-square statistic from the observed and expected frequencies we obtain a p-value, and if this value is less than a chosen significance level (often taken as 0.05), the null hypothesis is rejected. We note, that NA values in the data are excluded while performing this test.

%{\bf You may have to write the justification that the statistic under consideration would indeed be chi-squared distributed under the null hypothesis.}

\subsection{Teacher's Survey}
As part of the baseline survey, we have also interviewed the science teachers of the schools where the survey for the students was conducted to gain insights on the merits and challenges associated with the astronomy content and the pedagogical approach in schools. We will elaborate on the results of this qualitative study separately (Singhal et al, in prep).

% -----------------------------------------------
\section{Findings}
\label{sec:results}

After the data from the survey respondents is encoded and digitized, we analyze it statistically. Following the questions, the survey analysis is categorized into five domains: attitudes toward existing curricula, levels of interest in astronomy, knowledge of basic astronomy, cultural affiliations with astronomy, access to astronomical resources, and further interest in the subject (details in $\S$~\ref{sec:methods}). Furthermore, responses are divided into distinct groups based on school resource availability, distinguishing between resource-rich and resource-poor schools ($\S$~\ref{sec:methods}), as well as according to gender demographics, where applicable.

\subsection{Attitude toward current astronomy curricula}

\begin{figure}[h]
    \includegraphics[width=0.45\textwidth, trim = 0 80 0 0, clip]{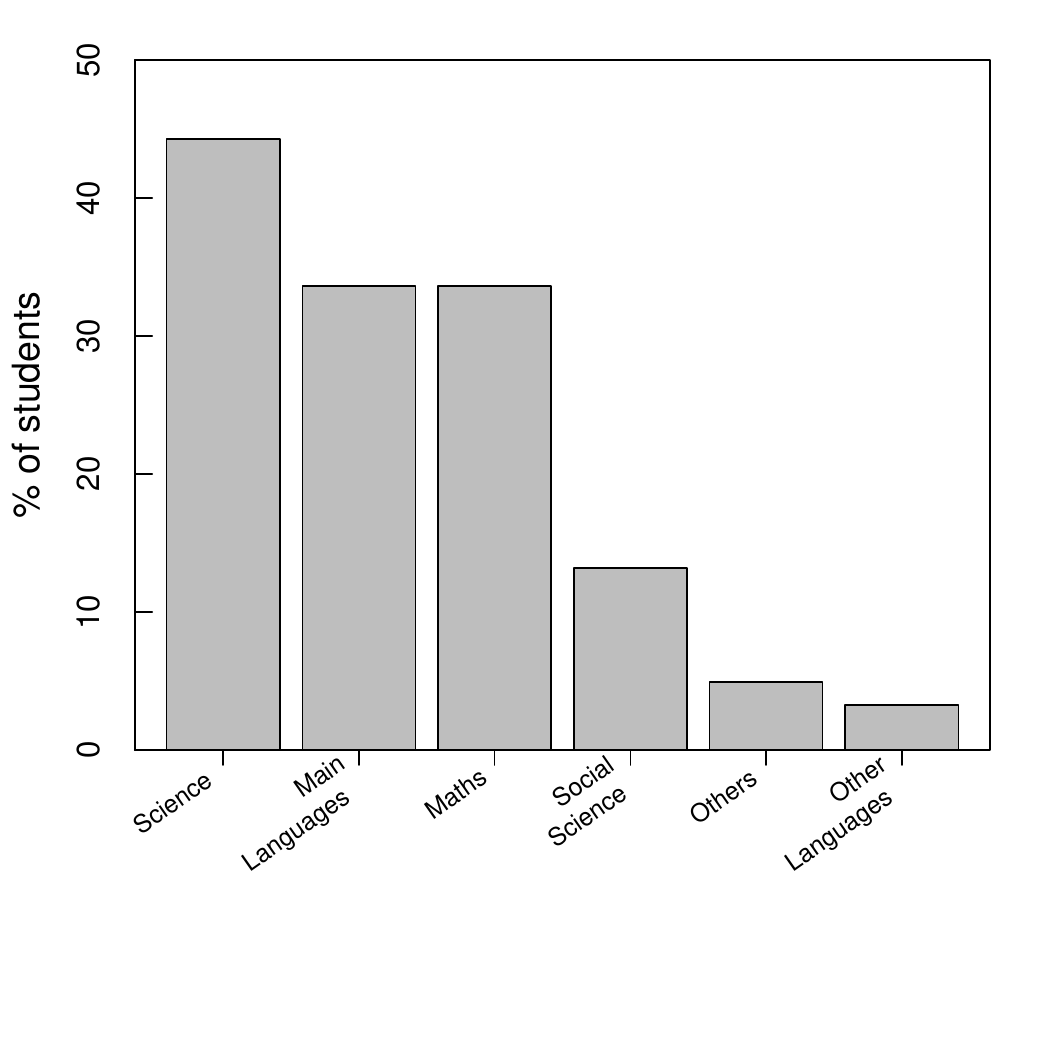}
    \caption{Favourite subject of the students. Many students gave multiple subjects as their favorite, this barplot takes into account all of the answers.}
    \label{fig:favsub}
\end{figure}

The first question in the survey asks the students to name their favorite subject. Figure~\ref{fig:favsub} shows the percentage of students that choose the different subjects as their favorite. Many students gave multiple subjects as their answers and in the analysis, we have accounted for all the subjects mentioned. Therefore, the percentages across the responses sum to more than hundred. While encoding the data from the pilot survey, some keywords emerged from the responses that we used to categorize the data from the main survey. The keywords used here are as follows: Maths (for all maths-related responses, e.g. geometry, algebra), Science (for all science-related entries, e.g. science, physics, chemistry, astronomy, biology), Social Studies (history, geography, civics, economics, psychology, etc), Main Languages (English / Hindi / Language of the respective state), Other Languages (e.g. Sanskrit, French, German, etc.), and Other (anything else, e.g. art, sports, etc). We find that science is the most popular subject, with almost $45\%$ of students favoring it.  Main languages and Maths share the second place with $34\%$ of students choosing each of them. This is followed by social science ($13\%$ of students), other subjects ($5\%$ of students), and other languages ($3\%$ of students).

\begin{figure*}[h]
    \centering
    \includegraphics[width=0.3\textwidth]{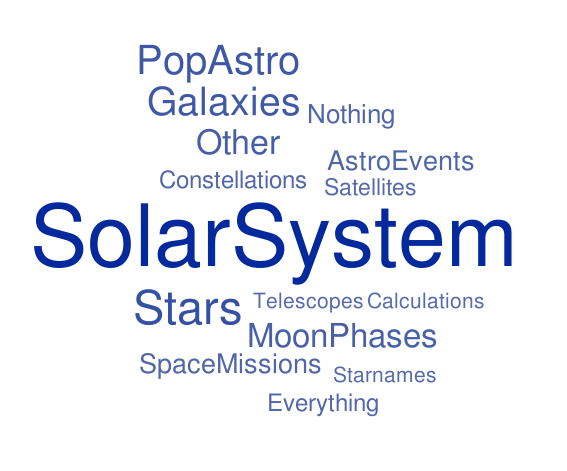}
    \includegraphics[width=0.3\textwidth]{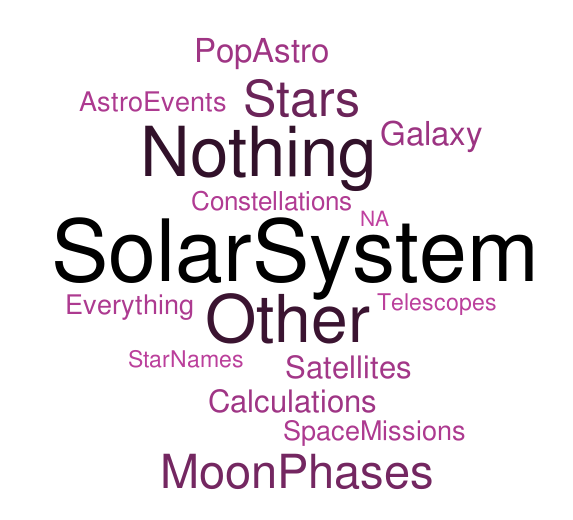}
    \includegraphics[width=0.3\textwidth]{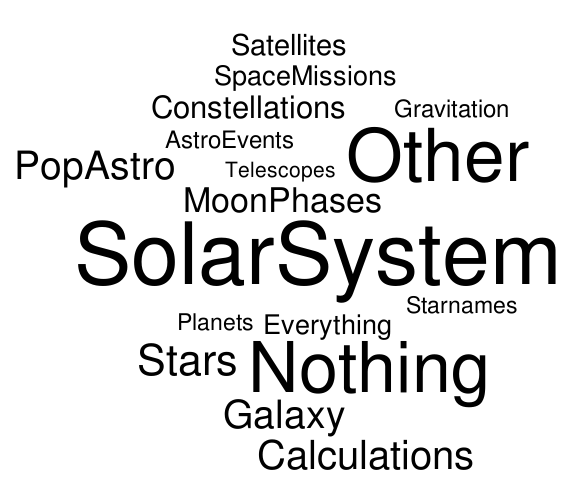}
    \caption{Word clouds representing the most interesting (left), least interesting (middle), and hardest (right) astronomy-related topics in the school curricula, as answered by the survey participants.}
    \label{fig:most_least_hardest}
\end{figure*}

Next, students were prompted to identify the astronomy topics that they found to be the most interesting, least interesting, and hardest in their curricula. Students gave a wide variety of responses to these, yielding substantial textual data that was systematically categorized based on trends that emerged from the pilot survey phase. Specifically, thematic consolidation was undertaken, such as grouping various solar system components (e.g., Sun, Moon, Earth, Mars, planets, asteroids) under the overarching category "Solar System". Similarly, related words were grouped in categories of stars, galaxies, constellations, space missions, and astronomical events (e.g. eclipse, seasons, tides, etc). Notably, even though the question asks for topics within the curriculum, some students referenced popular astronomy topics, e.g. black holes, dark matter, dark energy, Universe, etc; which were collectively categorized as "Popular Astronomy (PopAstro)". The refined dataset is visually represented through word clouds in Figure~\ref{fig:most_least_hardest}. We find that "Solar System" emerges as the predominant term across all three categories, probably because most of the content in the syllabus comes under this label. Additionally, students exhibit notable interest in stars, galaxies, and popular astronomy topics. On the other hand, stars and the phases of the Moon were recurrently cited as the least engaging astronomy topics. Furthermore, many students noted that they found calculations and stellar astronomy topics to be difficult. Among other recurrent hard topics were motions of the Earth and other celestial bodies (grouped under 'Other'). However, a large proportion of students ($11\%$ of students) commented that they found nothing in astronomy particularly hard or uninteresting.

%\textbf{Shall we report figures with NA included or not? What about the different groups and their comparison? Comparison to be updated.}
%\begin{figure}
%    \centering
%    \includegraphics[width=0.45\textwidth]{figures/highed_pie_data_v2.pdf}
%    \caption{Interest in learning astronomy in higher education}
%    \label{fig:enter-label}
%\end{figure}

\subsection{Knowledge of basic Astronomy}
To assess students' knowledge of basic astronomy, three questions were posed to gauge their understanding of sizes, distance scales, and moon phases.

\subsubsection{Sizes}
\begin{figure}
    \centering
    \includegraphics[width=0.45\textwidth, trim = 0 0 0 40, clip]{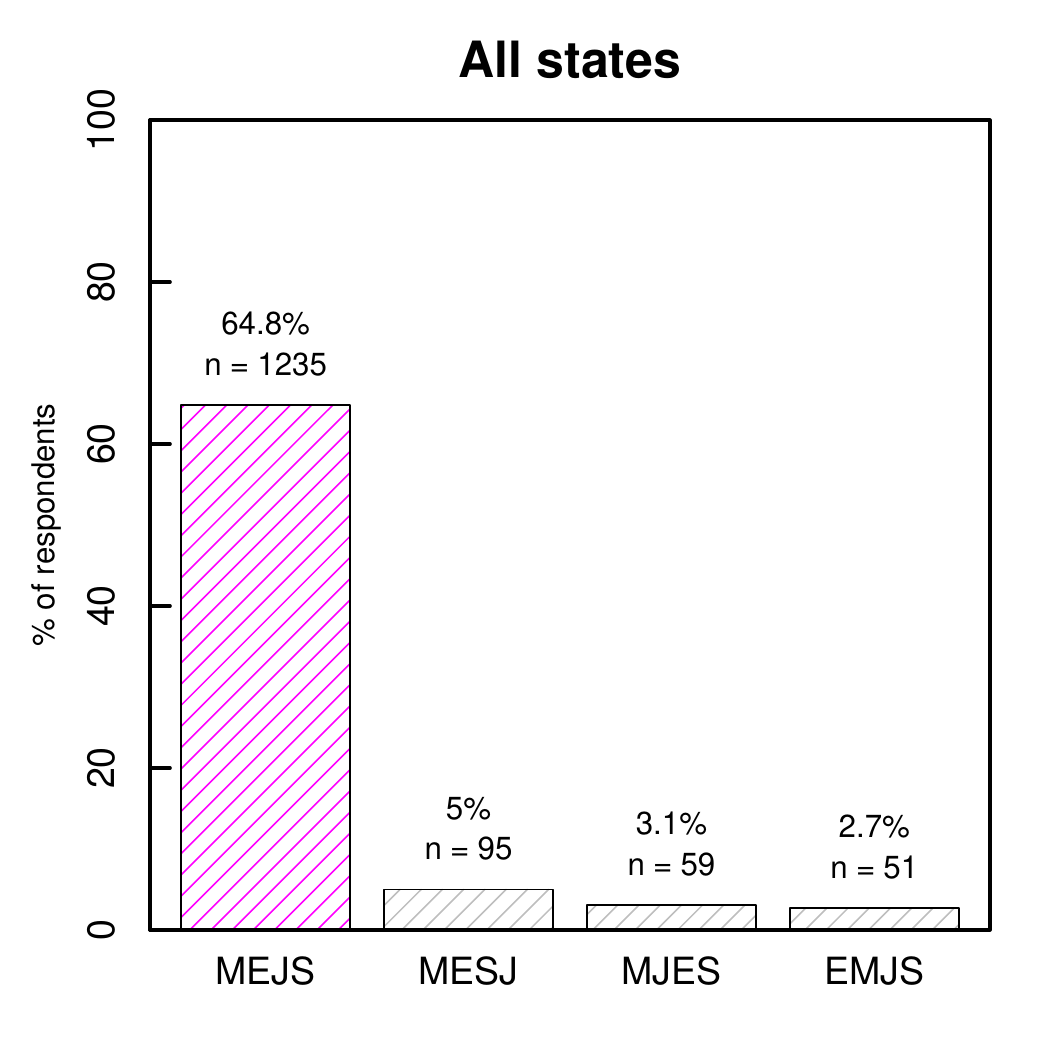}\\
    \caption{Bar chart showing the top four answers for the size-related question where students are asked to arrange Jupiter, Moon Earth, and Sun in order of size. The number and percentage of students who chose these options are also mentioned. Here E, M, J, and S are abbreviations used for Earth, Moon, Jupiter, and Sun respectively. The bar with magenta shading is the correct answer. }
    \label{fig:size}
\end{figure}

For the size-related question, students were tasked with arranging Jupiter, Moon, Earth, and Sun from smallest to largest. Figure~\ref{fig:size} shows the top four answers for this question. We find that $65\%$ of all students could correctly answer this question. There are no clear alternate conceptions regarding size rankings; the second, third, and fourth most common answers are given by $5\%, 3\%$ and $3\%$ people respectively. From the first top three answers, we find that the Moon (M), Earth (E), and Sun (S) are in the correct order, but students are uncertain about the placement of Jupiter in the sequence, as it was positioned in the third (correct option), fourth, and second place, respectively.

We further analyze the data of this question across different demographic groups. Among male respondents, a notable proportion, comprising $73\%$ of students, provided correct responses, whereas among female students $60\%$  answered the question accurately.  On the other hand, $85\%$ of students from resource-rich schools answered it correctly in contrast to only $50\%$ of resource-poor students. The disparities in response accuracy are highly statistically significant across both gender lines (p-value < \num{e-7}) and resource distribution (p-value < \num{e-15}) categories.

\begin{figure}
    \centering
\includegraphics[width=0.5\textwidth, trim= 0 0 0 40]{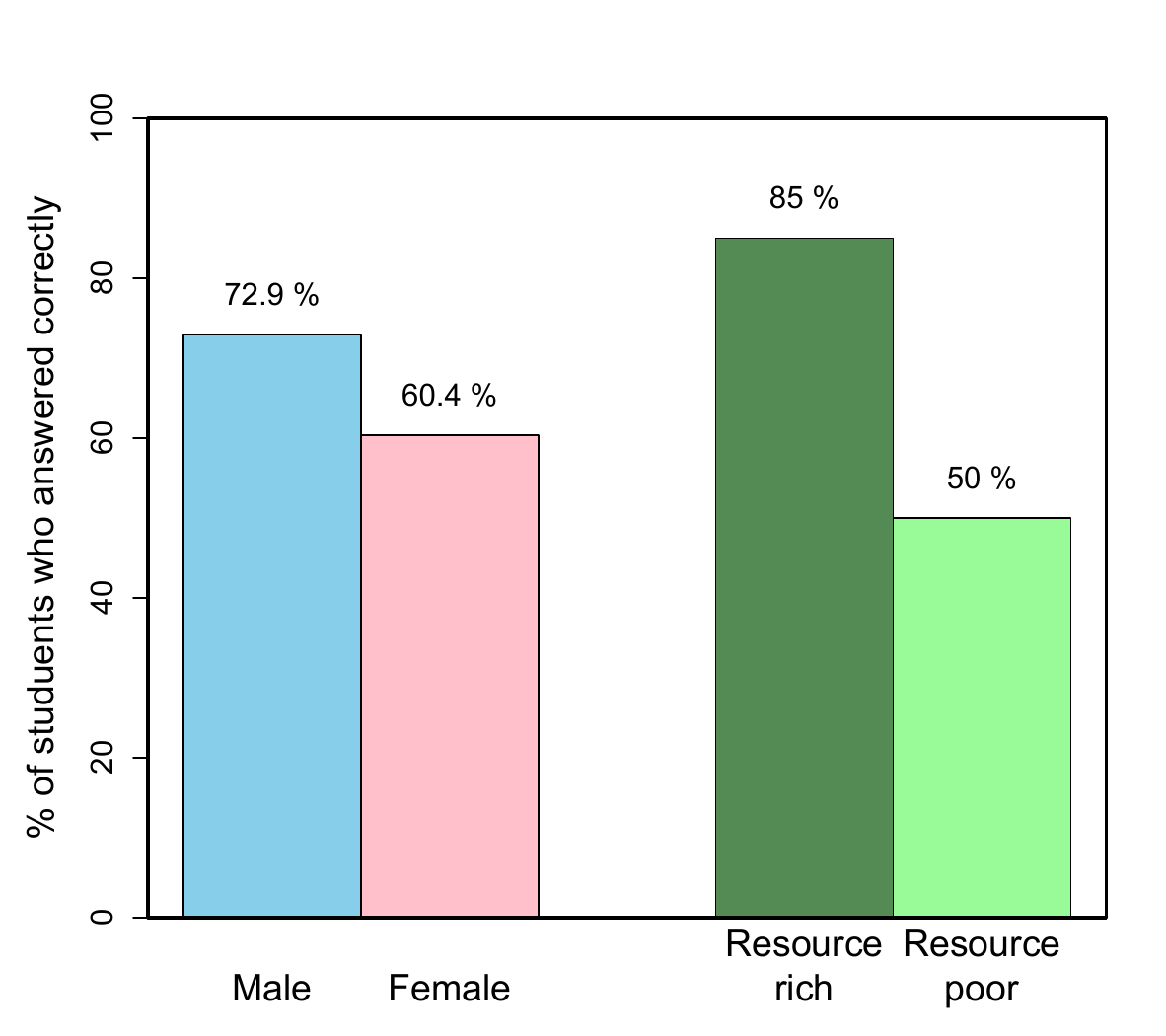}
    \caption{Comparison of percentage of students who answered the size-related question correctly across various demographic groups, including male and female students, as well as resource-rich and resource-poor students.}
    \label{fig:top_size}
\end{figure}

%\begin{figure*}
%    \centering
 %   \includegraphics[width=0.4\textwidth]{figures/size/size_bar_data_v2_percent_female.pdf}
 %   \includegraphics[width=0.4\textwidth]{figures/size/size_bar_data_v2_percent_male.pdf}
 %   \includegraphics[width=0.4\textwidth]{figures/size/size_bar_data_v2_percent_rich.pdf}
 %   \includegraphics[width=0.4\textwidth]{figures/size/size_bar_data_v2_percent_poor.pdf}    
 %   \caption{Same as Figure 4, but for female students (top left), male students (top right), students from resource-rich schools (bottom left) and resource-poor (bottom right) schools respectively.}
 %   \label{fig:enter-label}
%\end{figure*}

\subsubsection{Distances}
\begin{figure}[h]
    \centering
    \includegraphics[width=0.45\textwidth]{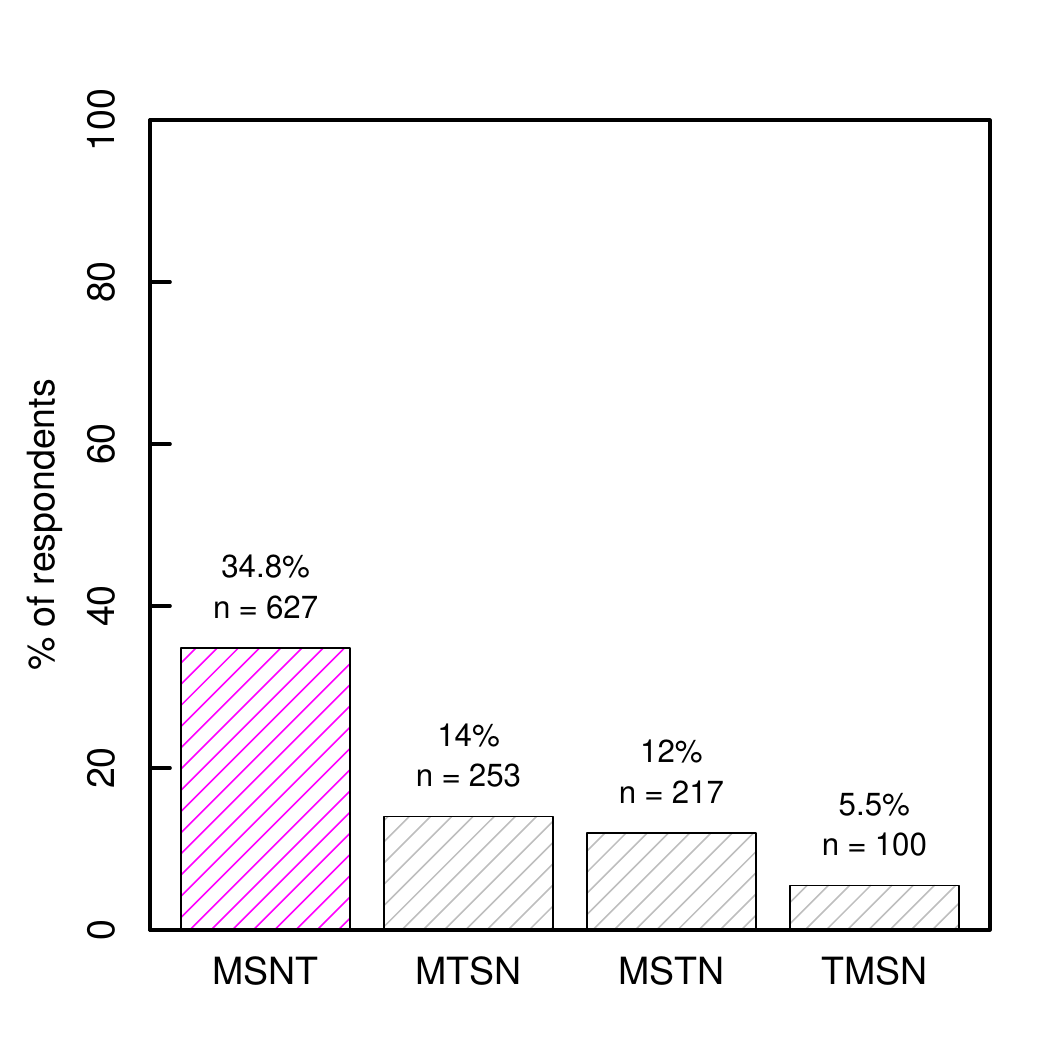}\\
    \caption{Bar chart showing the top four answers for Question 4 (arrange Sun, Moon, Stars, and Neptune in order of increasing distance from the earth) along with the number and percentage of students for each of them. Here S, M, T, and N are abbreviations used for Sun, Moon Stars, and Neptune respectively.}
    \label{fig:dist}
\end{figure}

In another question, students were asked to arrange the Sun, Moon, Stars, and Neptune in order of proximity (nearest to farthest) from the Earth. Figure~\ref{fig:dist} illustrates the top four choices among students. We found that only $35\%$ of students could answer this question correctly. Looking at the top choices we note that students can arrange Sun, Moon, and Neptune accurately, but they are unsure about the placement of stars. In the top four responses, stars were positioned as the farthest object (the correct choice) in the fourth position, and alternately in the second, third, and first positions (nearest object), respectively. These findings underscore a deficiency in students' comprehension of the astronomical distance scale, particularly about the expansive distances between stars compared to distances within the solar system.

\begin{figure}[h]
    \centering
    \includegraphics[width=0.5\textwidth]{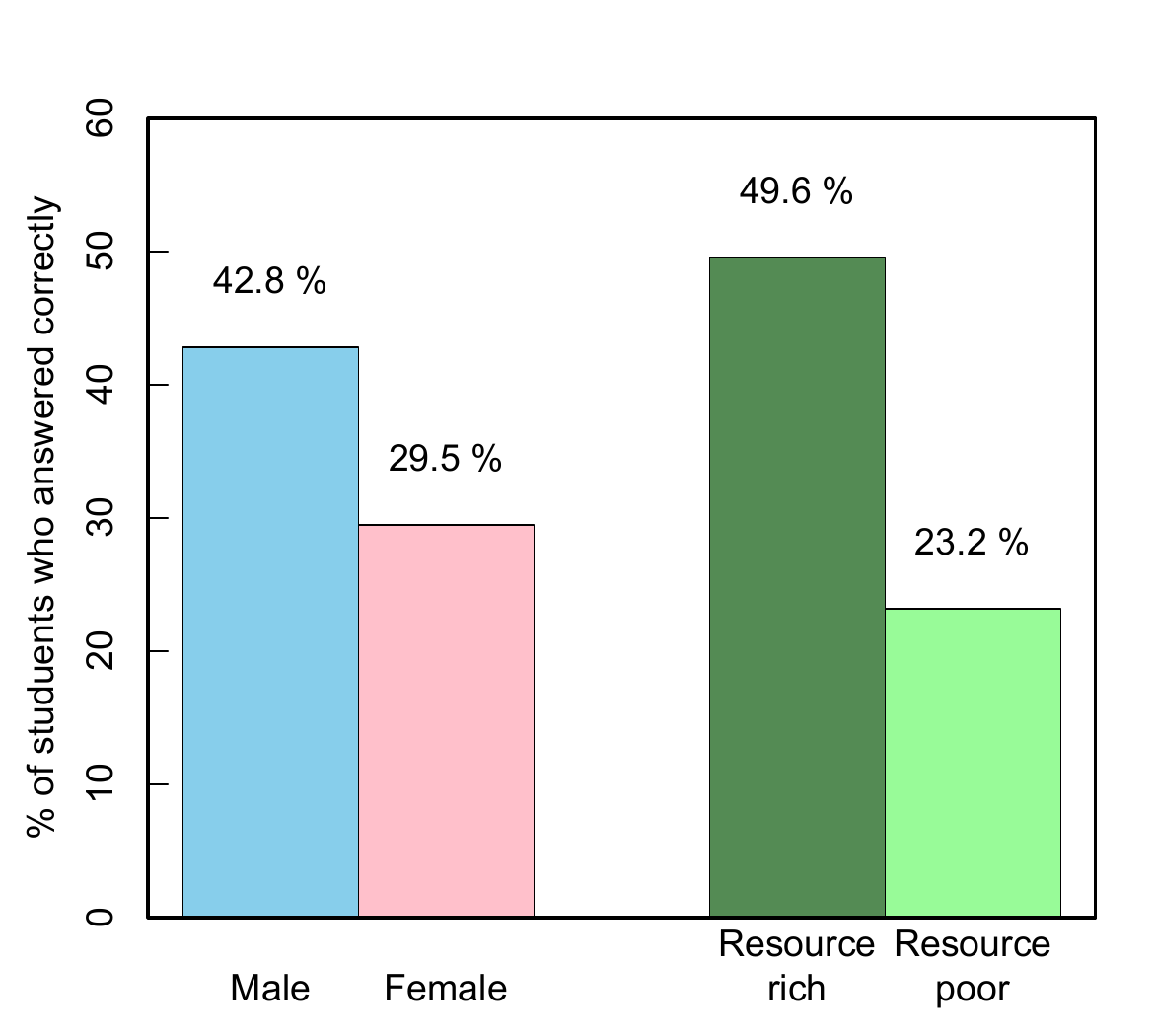}
    \caption{Comparison of percentage of students who answered the distance-related question correctly across various demographic groups, including male and female students and resource-rich and resource-poor students.}
    \label{fig:dist_all_groups}
\end{figure}

We further analyze this data along different demographic lines as shown in Figure~\ref{fig:dist_all_groups}. We find that among male students, $43\%$ provided a correct response, compared to $30\%$ of female students. Applying the chi-square test we find that the difference between genders is statistically significant with a p-value less than \num{e-7}. Along the socioeconomic lines,  $50\%$ of students from resource-rich schools answered the distance question correctly, whereas from resource-poor schools only $23\%$ of students were able to do so, again this difference is also highly statistically significant (p-value < \num{e-15}).  Overall, these findings indicate that irrespective of gender or socioeconomic background, a deficient understanding of distance scales persists among all student cohorts, although, the performance of male students and students from resource-rich schools is comparatively better than their counterparts.

\subsubsection{Phases of the Moon}
\begin{figure}[h]
    \centering
    \includegraphics[width=0.45\textwidth]{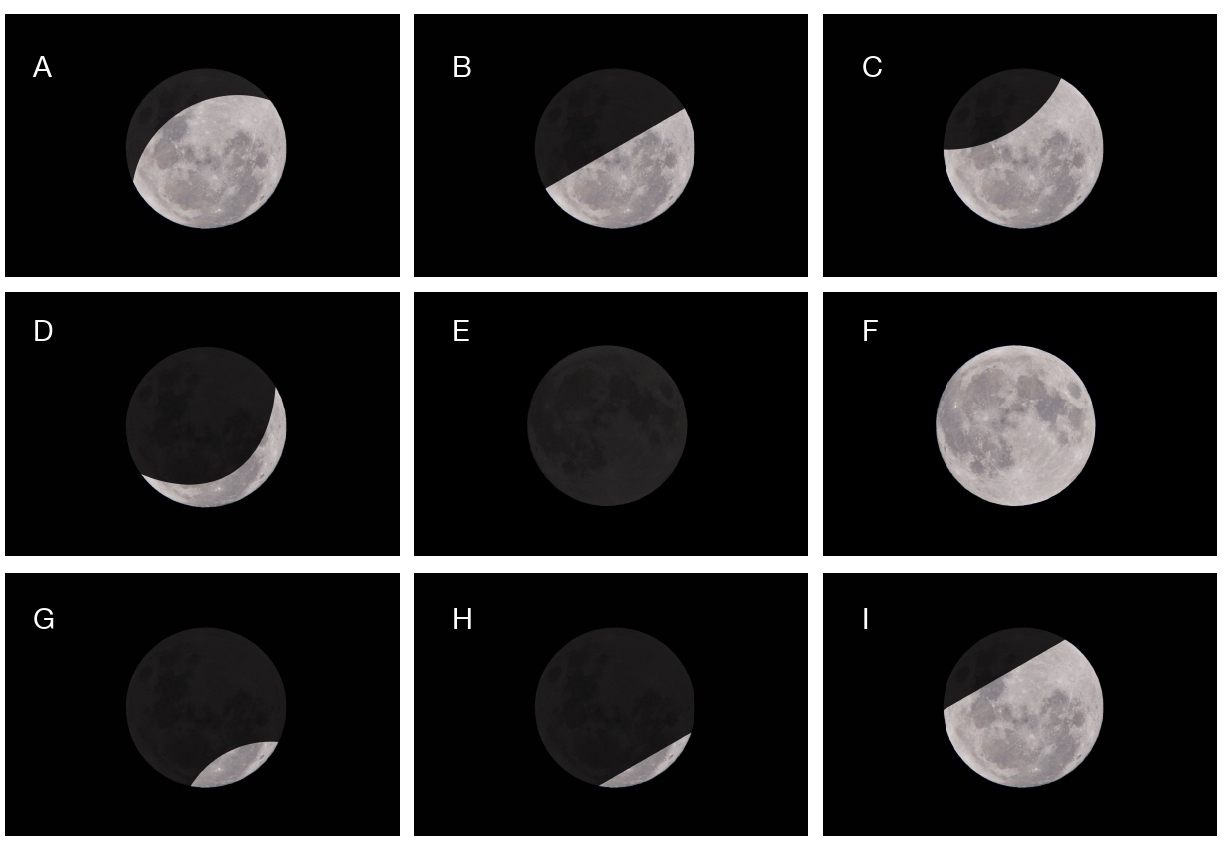}\\
    \caption{Students were given this set of nine images as possible choices and asked to select the correct ones for the following days in a lunar cycle: New Moon or Day 0, Day 2, Day 8, Day 11, and Full Moon.}
    \label{fig:moonimages}
\end{figure}

The next question was designed to test for basic knowledge about the phases of the Moon, not its causes, but just its appearance. Phases of the Moon are quite easy to observe even with the naked eye and are often suggested as an observational exercise for students in the textbooks. For this question, students were given a set of nine images (Figure~\ref{fig:moonimages}) and then they were prompted to choose the correct image for the following days in a lunar cycle: New Moon\footnote{The questionnaire mentions the new moon day as day 1, but conventionally, the new moon day is taken as day 0. In the text, we fall back to the conventional designation of the new moon day as Day 0 (The questionnaire in the appendix is retained as original).} or Day 0, Day 2, Day 8, Day 11, and Full Moon. We also mention the vernacular names for New Moon and Full Moon Day in the survey question to reduce the chance of confusion.

\begin{figure*}
\centering
    \includegraphics[width=0.33\textwidth]{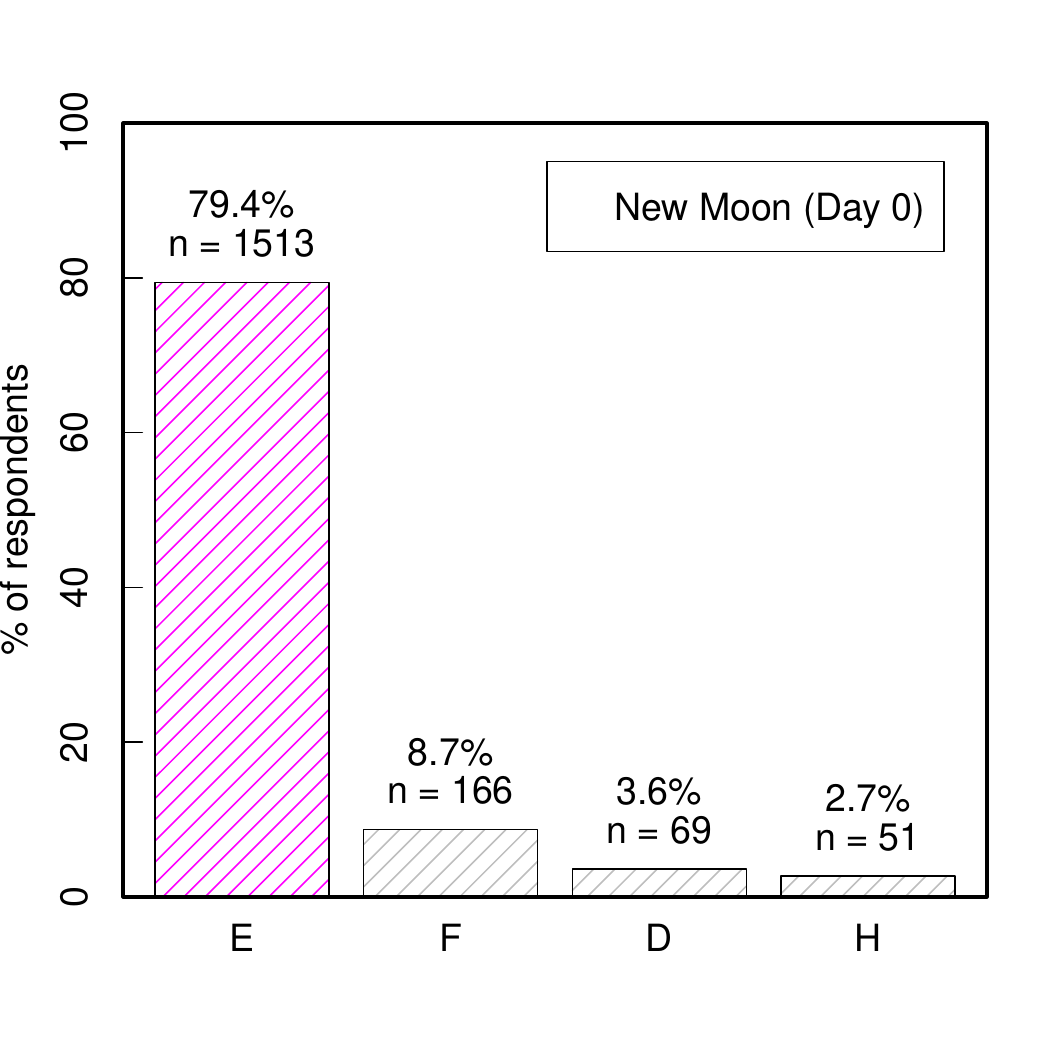}
    \includegraphics[width=0.33\textwidth]{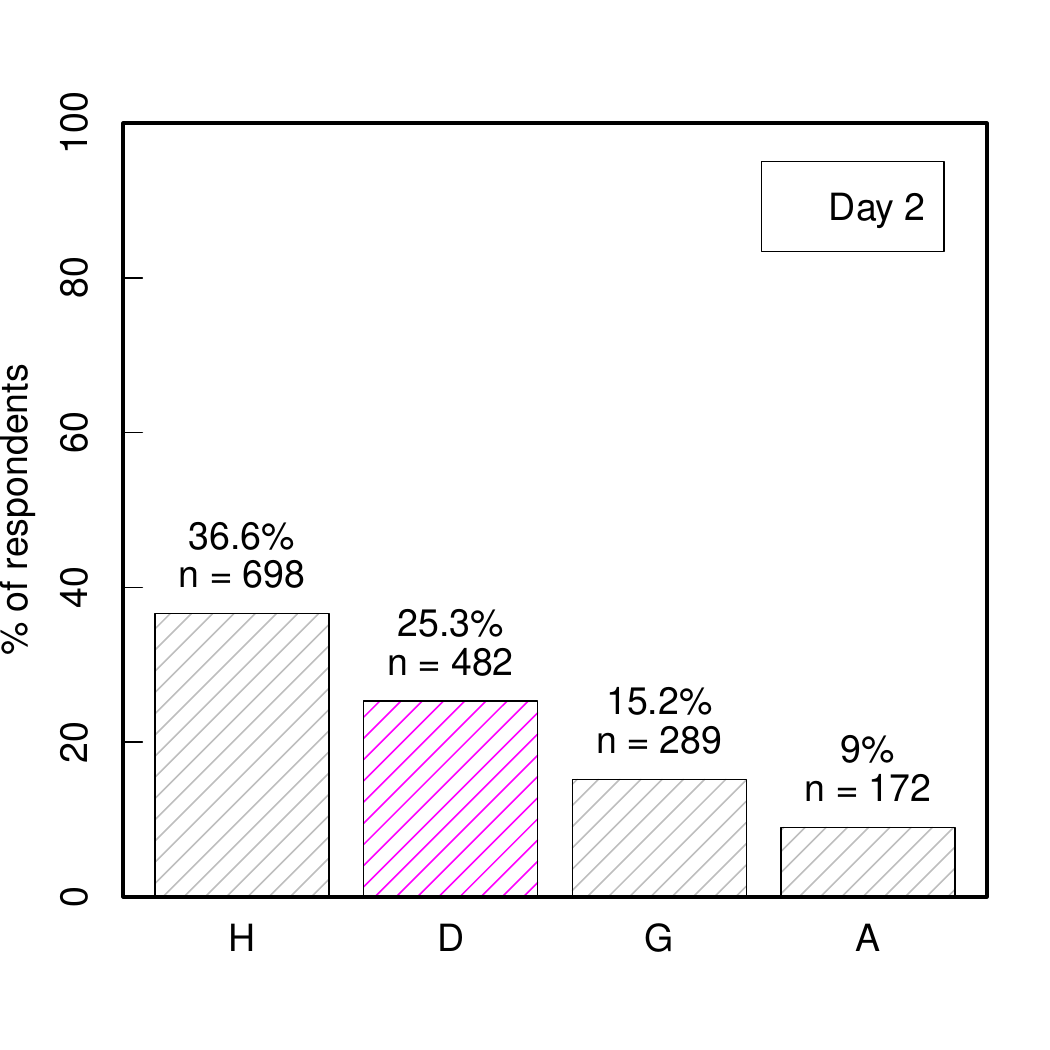}
    \includegraphics[width=0.33\textwidth]{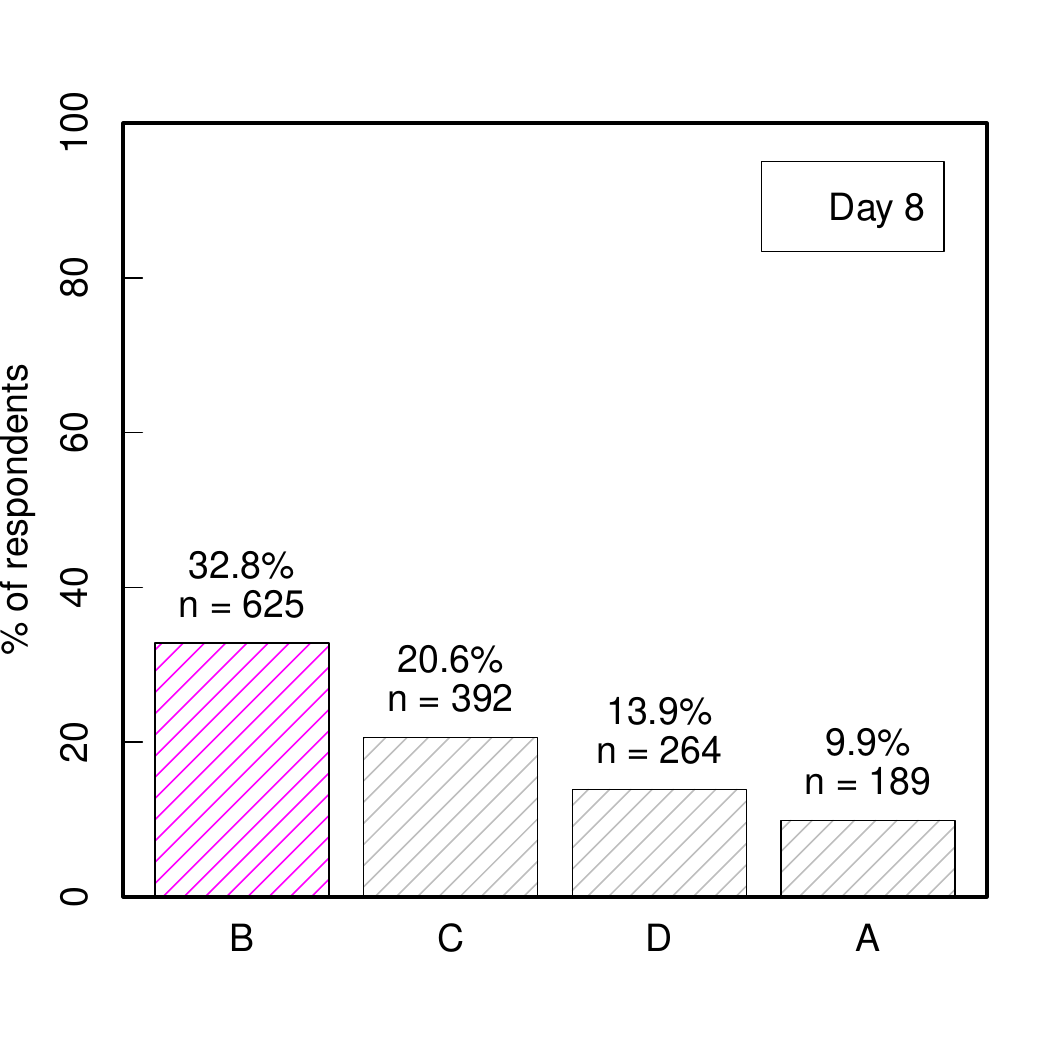}\\
    \includegraphics[width=0.33\textwidth]{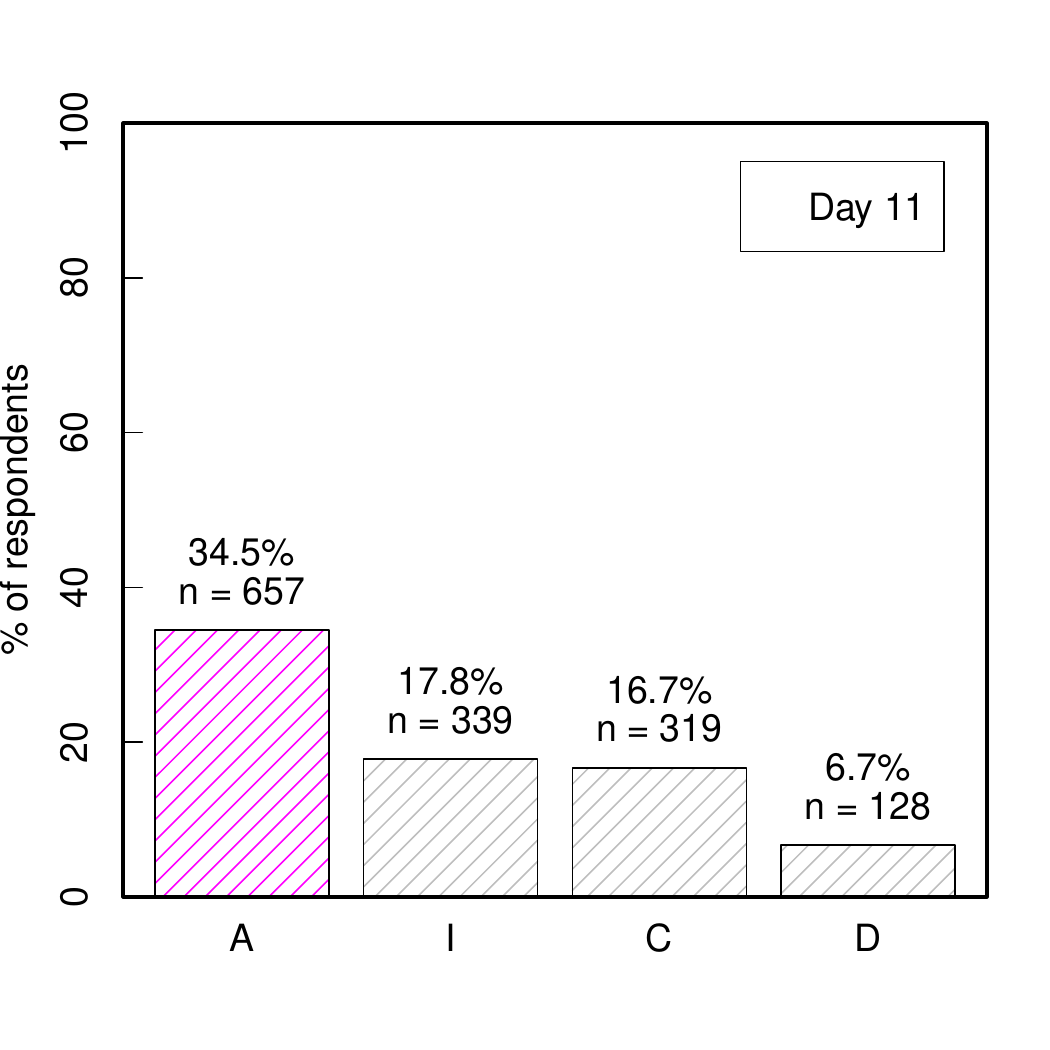}
    \includegraphics[width=0.33\textwidth]{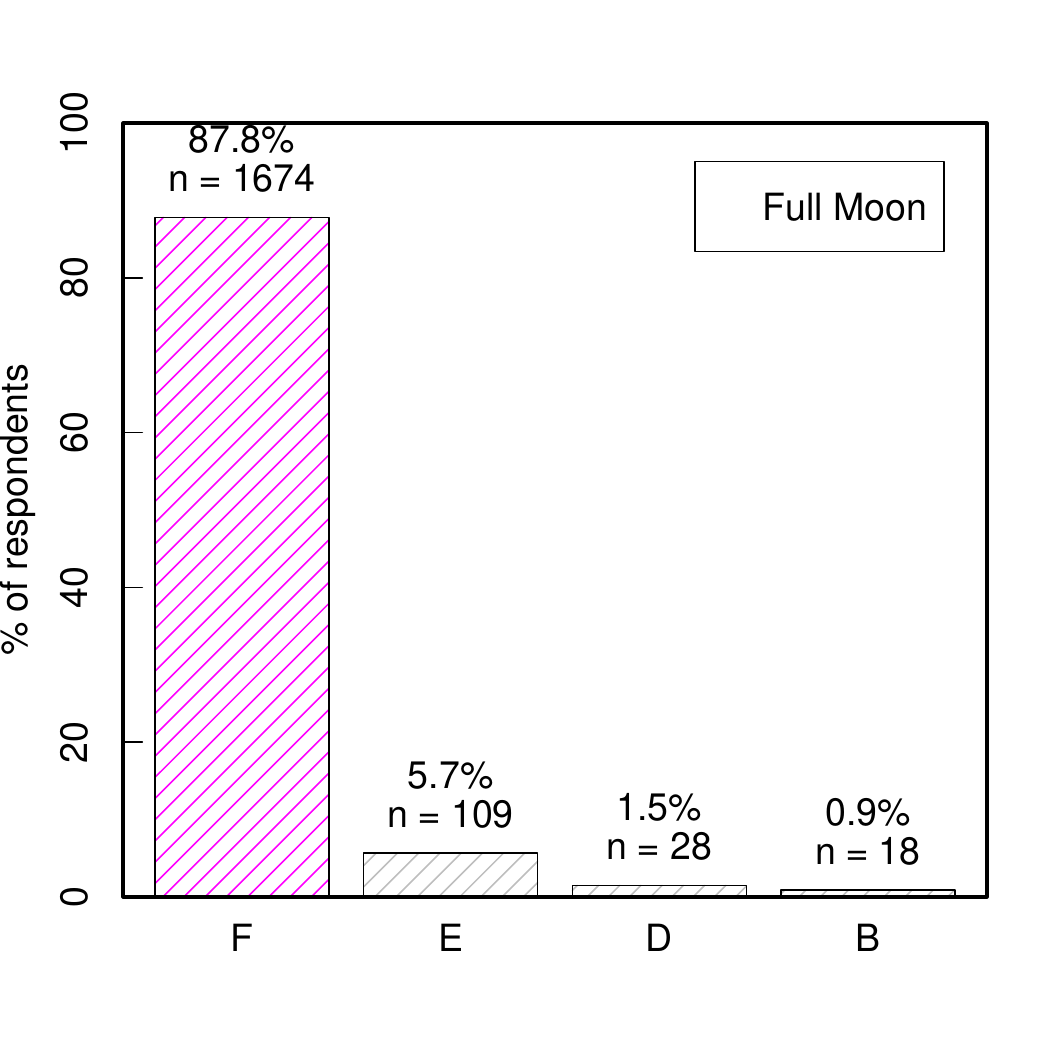}
     \caption{Histograms of numbers and percentage of students choosing the corresponding images for different days in a lunar cycle from a given set of nine images (Figure~\ref{fig:moonimages}).  The figures show the top four choices for New Moon or Day 0 (top left), Day 2 (top center), Day 8 (top right), Day 11 (bottom left), and Full Moon (bottom right). The correct answers for these five phases are marked in magenta shading.}
    \label{fig:phases}
\end{figure*}

The top four selections for each of these five lunar phases are presented in Figure~\ref{fig:phases}. We find that $79\%$ of all students correctly identified Image E (from Figure~\ref{fig:moonimages}) as depicting the New Moon phase. The second most popular choice ($9\%$ of students), surprisingly, corresponds to Image F which actually depicts a Full Moon. This discrepancy may stem from a simple misunderstanding of the question or an uncertainty in the nomenclature, which seems to persist even after the regional names were clarified in the question. It would be worth noting here that several indigenous communities refer to the full moon as ``bright as new'' in their languages \citep{Shetye2023}. Thus, it may be natural for some students to associate word ``new'' with the full moon. In the case of Full Moon, we find that $88\%$ of students identified the correct image (image F). However, the second most popular choice for this phase, selected by $6\%$ of students, was Image E (corresponding to the new moon), likely indicative of a similar misunderstanding. Nonetheless, the vast majority of students demonstrated proficiency in correctly identifying images depicting both the New Moon and Full Moon phases.

However, a distinct pattern emerges for the remaining three lunar phases. For Day 2, only $25\%$ of students correctly identified Image D as depicting the lunar phase, with the correct answer being the second most popular choice rather than the first. Image H, which has a straight line shaped terminator line (line separating day and night on the Moon), garnered the highest frequency of selections, chosen by $37\%$ of students, while Image G, which also has an incorrect terminator line, ranked third in popularity. The prevalence of these top three choices indicates that while students may infer that the Moon would appear small on Day 2, there is a considerable uncertainty regarding its shape.

For Day 8, the correct selection (Image B) was chosen by $33\%$ of students, making it the most popular choice. However, Image C (21$\%$ of students), Image D (14$\%$ of students), and Image A (10$\%$ of students) also received notable frequencies of selection. We note that in order to correctly identify the image corresponding to Day 8 in the lunar cycle, students must possess an understanding of the duration of the lunar cycle and deduce that Day 8 falls midway between the New Moon and Full Moon phases. The prevalence of uncertainty among the top choices suggests that many students lack clarity on this aspect.

For Day 11, Image A (correct choice) emerged as the most popular choice, selected by $35\%$ of students, followed by Images I ($18\%$ of students), C ($17\%$ of students), and D ($7\%$ of students). Once again, ambiguity regarding the moon's shape is evident among the top three choices, with uncertainty prevailing regarding whether the moon exhibits a concave, straight, or convex edge.

These findings underscore a widespread confusion among students in simply recognizing the shapes of different moon phases. It is indicative that a significant proportion of students may not have critically observed lunar phases, despite it being a prescribed activity in nearly all textbooks. Furthermore, it suggests a potential lack of understanding regarding the underlying mechanisms driving Moon phases. Notably, some of the top selections imply a mental model for the Moon phases akin to an eclipse, thereby explaining the prevalence of responses depicting particular shapes for the terminator line.

\begin{figure*}[h]
\centering
    \includegraphics[width=0.33\textwidth]{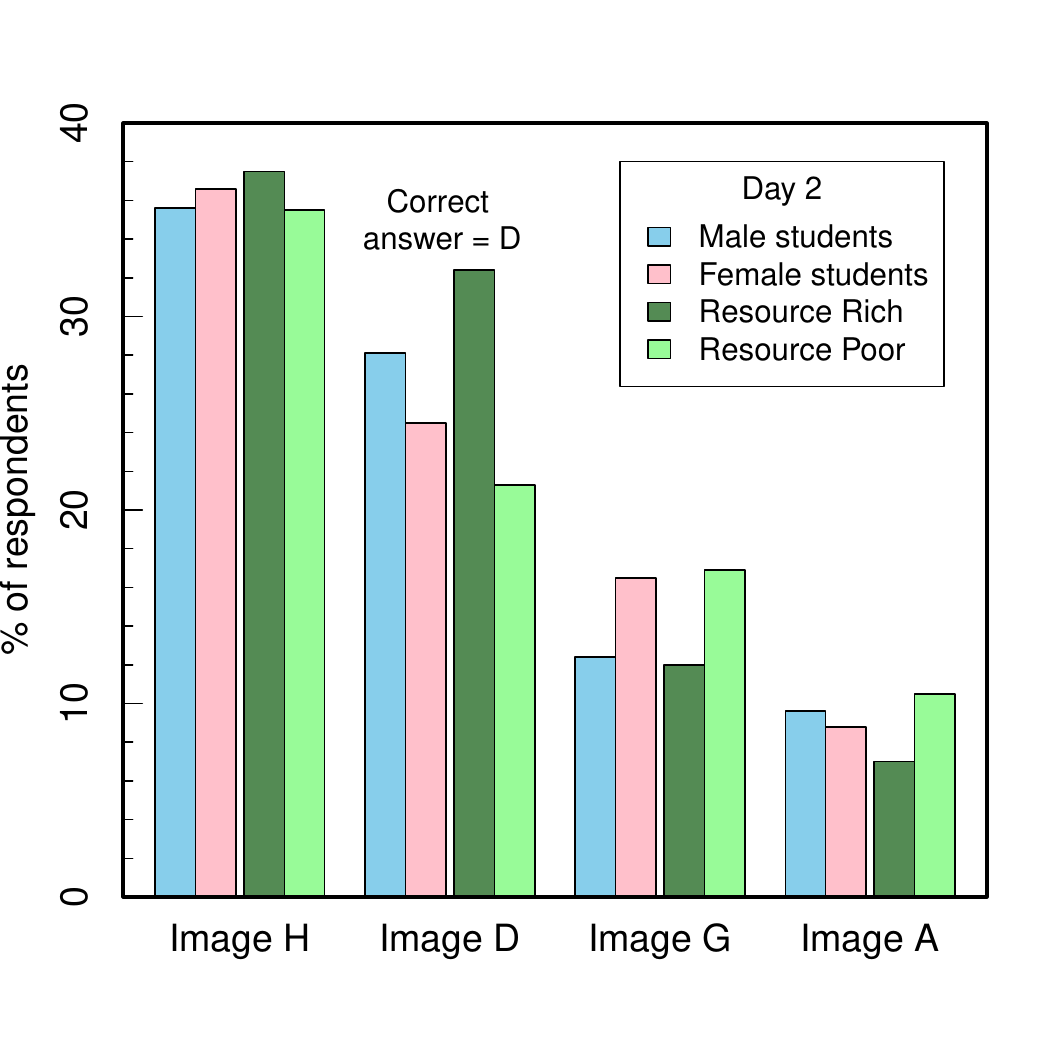}
    \includegraphics[width=0.33\textwidth]{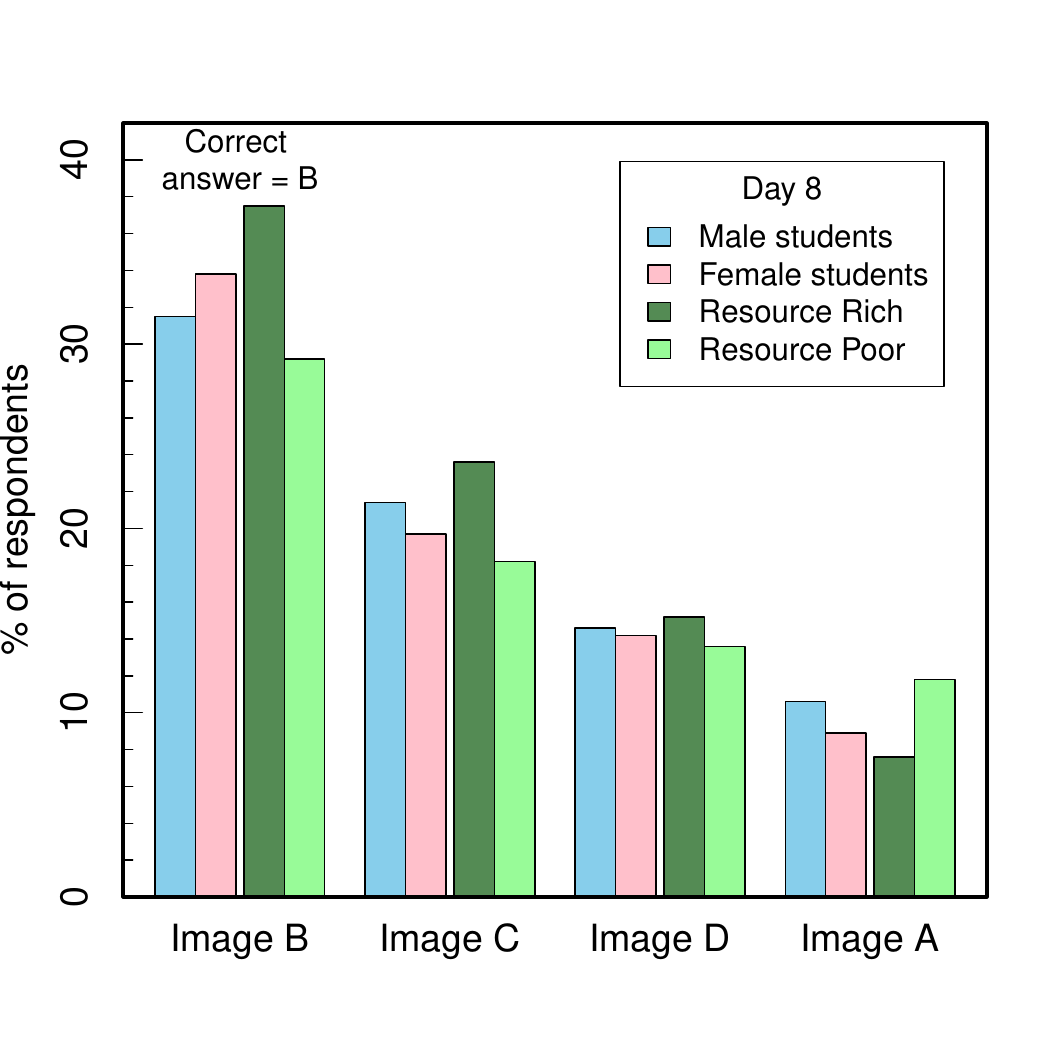}
    \includegraphics[width=0.33\textwidth]{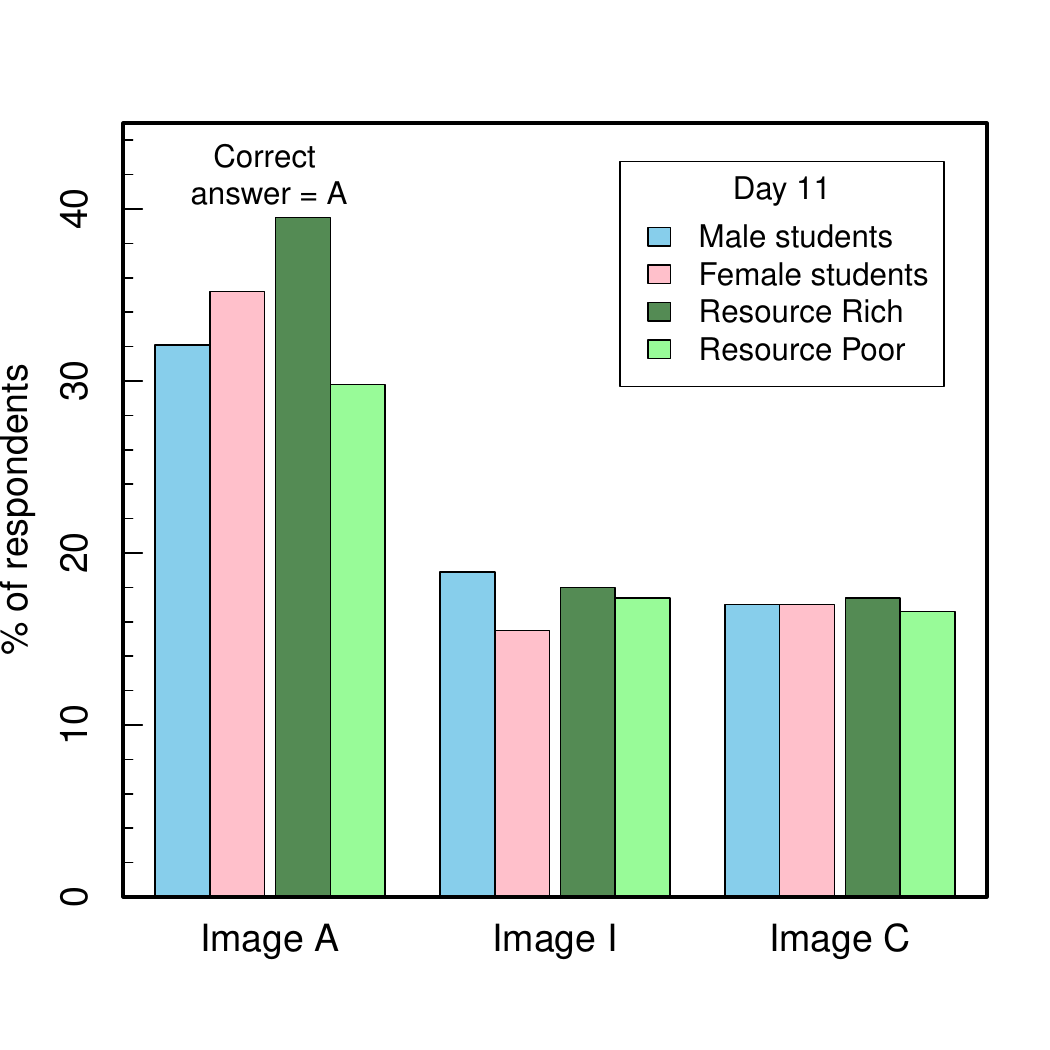}
    \caption{Histograms of percentages of students choosing the corresponding figures for different Moon Phases, Day 2 (left), Day 8 (middle), and Day 11 (right) for the different demographics, male (blue) and female (red) students, along with students from resource-rich (dark green) and resource-poor (light green) schools. The correct options are noted in the plots.}
    \label{fig:phases_demo}
\end{figure*}

We further analyze this data along the different demographic divisions. Figure~\ref{fig:phases_demo} shows the top choices for Days 2, 8, and 11 for male and female students as well as those from resource-rich and resource-poor schools. We find that the top answers are consistently the same among all groups. Upon closer examination of the correct responses, we observe that for Day 2, 8 and 11 female and male students perform almost similarly ( the differences are statistically insignificant with p-value = 0.1, 0.33, and 0.19 for days 2, 8, and 11 respectively). Additionally, students from resource-rich schools perform better than their resource-poor counterparts, and in this case, the difference is statistically significant (p-value < \num{e-06}, \num{e-3}, and \num{e-4} for days 2, 8, and 11 respectively). Similar trends were present for the New Moon and Full Moon phases also. We discuss further implications of these results in section \ref{sec:discussion}.

Overall, we find that only $7\%$  of all students have correctly identified images for all five phases. There was statistically no gender difference in the percentages of students answering the question correctly. Additionally, the resource-rich school's students demonstrated relatively higher proficiency at $9\%$ compared to those from resource-poor backgrounds, who achieved a lower rate of $4\%$. The fraction is dismally low among all demographics which suggests a severe lack of understanding of moon phases among students.

\subsection{Cultural Connections with Astronomy }

\begin{figure*}[h]
    \centering
    \includegraphics[width=0.45\textwidth]{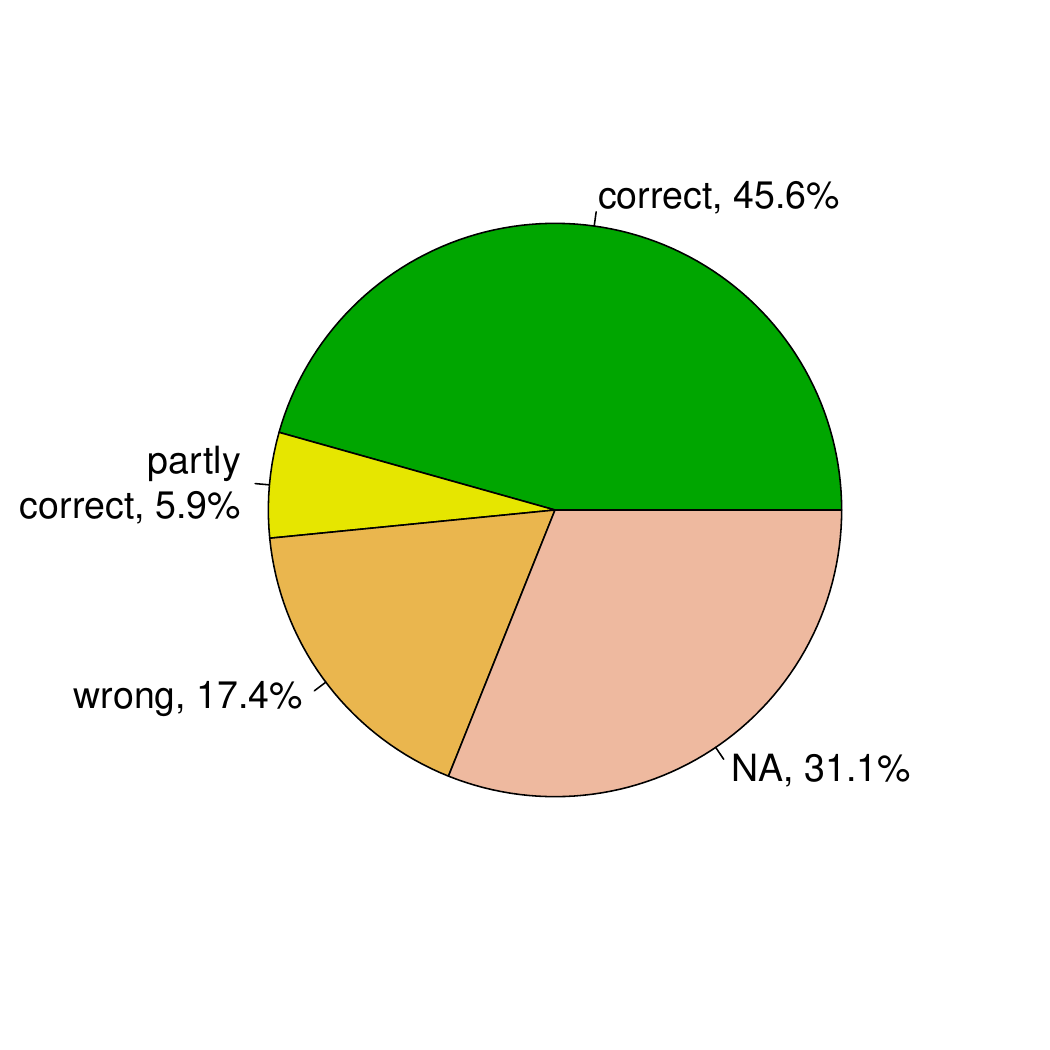}
    \includegraphics[width=0.45\textwidth]{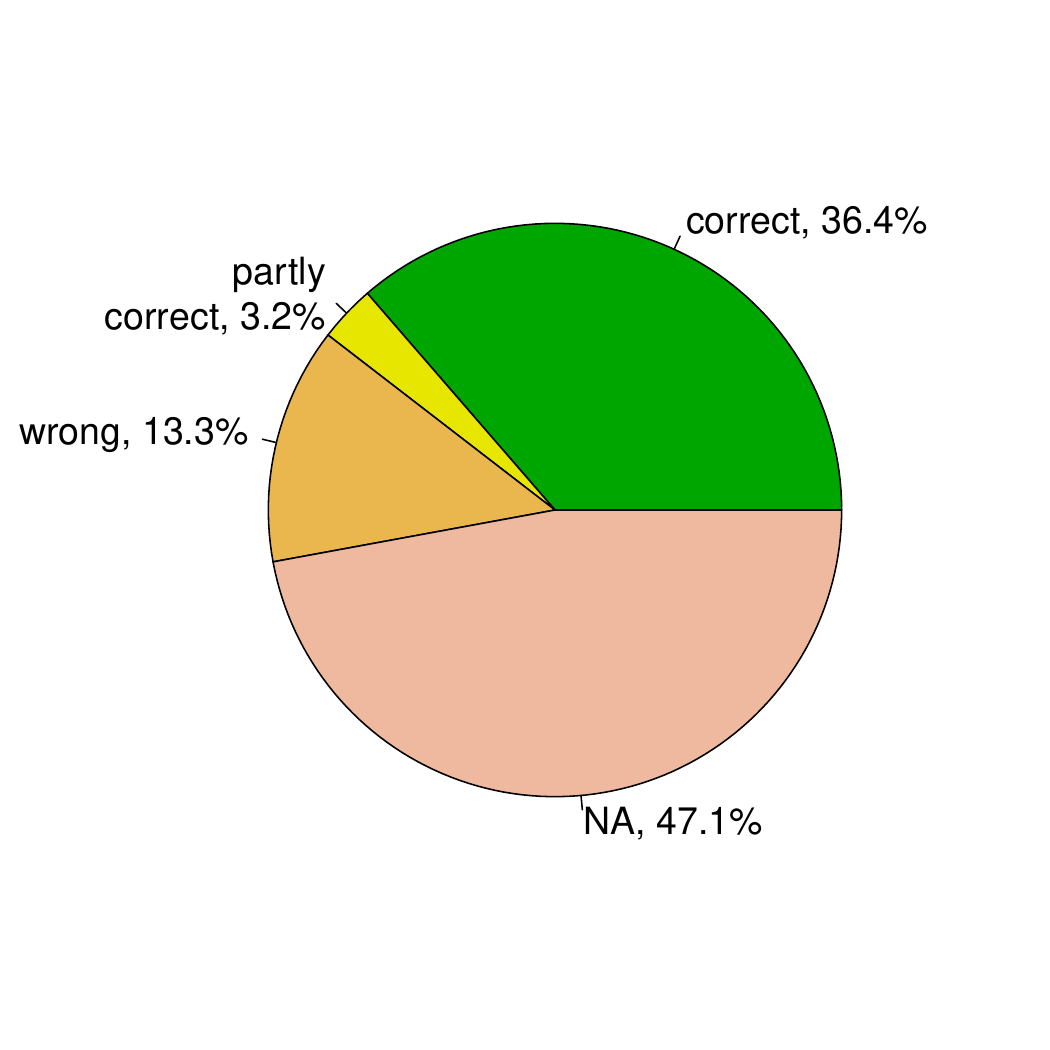}    
    \caption{Pie chart of student responses for naming some festivals that always fall on Full Moon day (left) and New Moon day (right). The responses are divided into three categories: correct, partially correct, and wrong. Non-response is indicated as `NA'.}
    \label{fig:fest}
\end{figure*}

India boasts a rich historical legacy of astronomical inquiry, with many of its cultural practices and festivals closely aligned with astronomical phenomena. The majority of Indian festivals adhere to the lunar calendar, underscoring the intrinsic connection between astronomy and cultural traditions. Recognizing the influence of astronomy in everyday life, our study sought to ascertain the extent to which students are cognizant of this cultural linkage. To explore this, the students were asked to name some festivals that are celebrated on the Full Moon day and New Moon day respectively. The distribution of student responses is depicted in Figure~\ref{fig:fest}. We find that for Full Moon festivals, $46\%$ of students answered it correctly, with a further $6\%$ answers being partially correct (i.e. among the festivals they mentioned for this answer, some were wrong). Conversely, a notable portion, comprising $17\%$ students provided wrong answers, and $31\%$ students refrained from answering this question. On the other hand, for New Moon, almost half of the students ($47\%$) did not answer this question at all, and only $36\%$ of all students could answer it correctly. These results underscore a gap in student's awareness of the cultural connection of these festivals to astronomy.

\begin{figure}[h]
    \centering
    \includegraphics[width=0.45\textwidth, trim= 0 0 0 30]{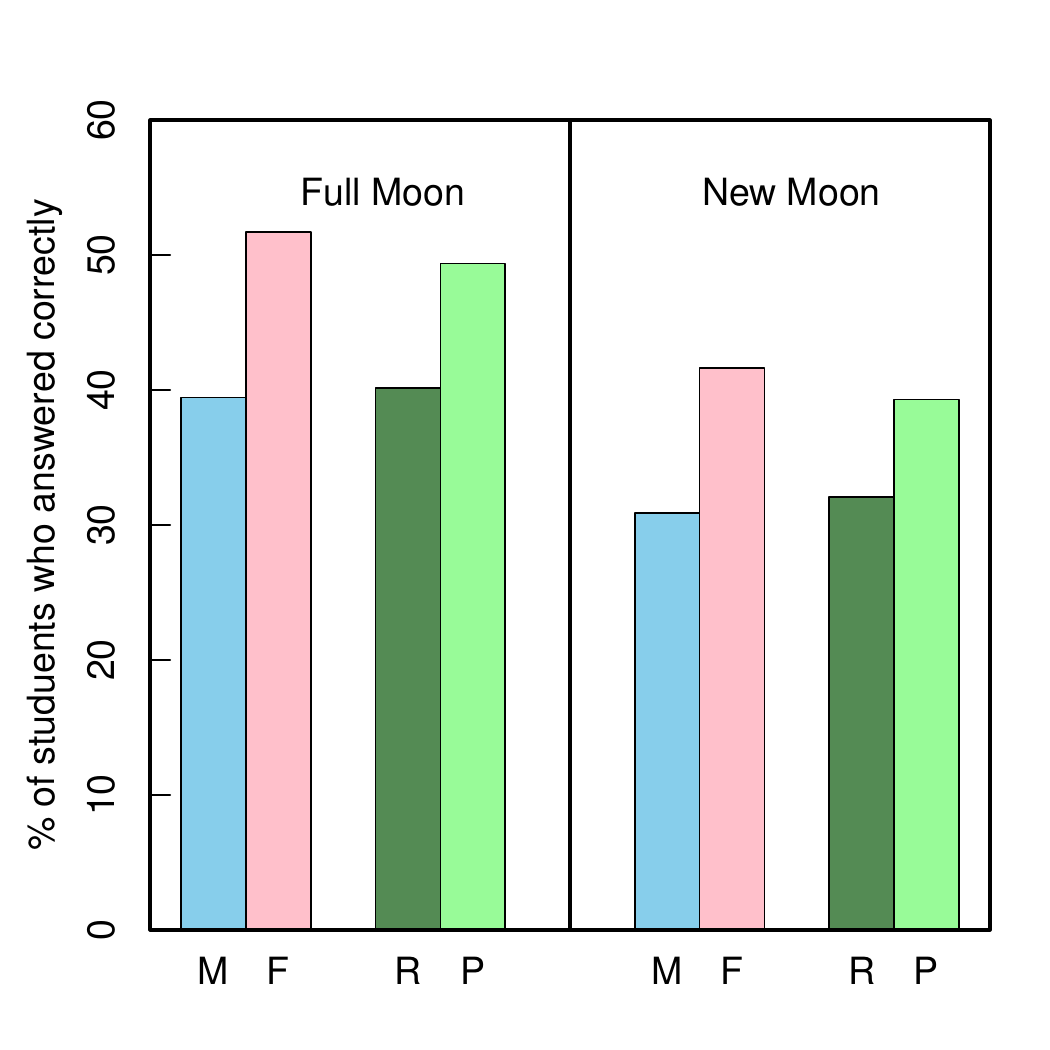}
    \caption{Comparison of percentage of students who correctly answered for festivals celebrated on Full Moon days (left) and New Moon days (right) across various demographic groups. Here M, F, R, and P are abbreviations for Male, Female,  Resource-rich school students, and Resource-poor school students respectively.}
    \label{fig:fest_demo}
\end{figure}

Similar to before, we perform further analysis on this data by dividing it into different demographic groups (Figure~\ref{fig:fest_demo}) and find that for both phases, female students outperform males, and students from resource-poor schools perform much better than their resource-rich counterparts. These differences are statistically significant, highlighting the pronounced disparities across demographic lines in terms of cultural awareness linked to astronomical traditions. Although the reasons behind this are not explicitly elicited in any follow-up with these groups, it may be speculated that groups such a resource rich students or male students are socially pressured to spend more time with ``formal studies'', i.e. textbooks, assignments and other co-curricular materials and are discouraged from spending time on supposedly ``lesser pursuits'' like observing the nature or listening to stories from elders.

\subsubsection{Belief in Astrology}

\begin{figure}[h]
    \centering
    \vspace{0.1in}
    \includegraphics[width=0.47\textwidth, trim= 0 20 0 50]{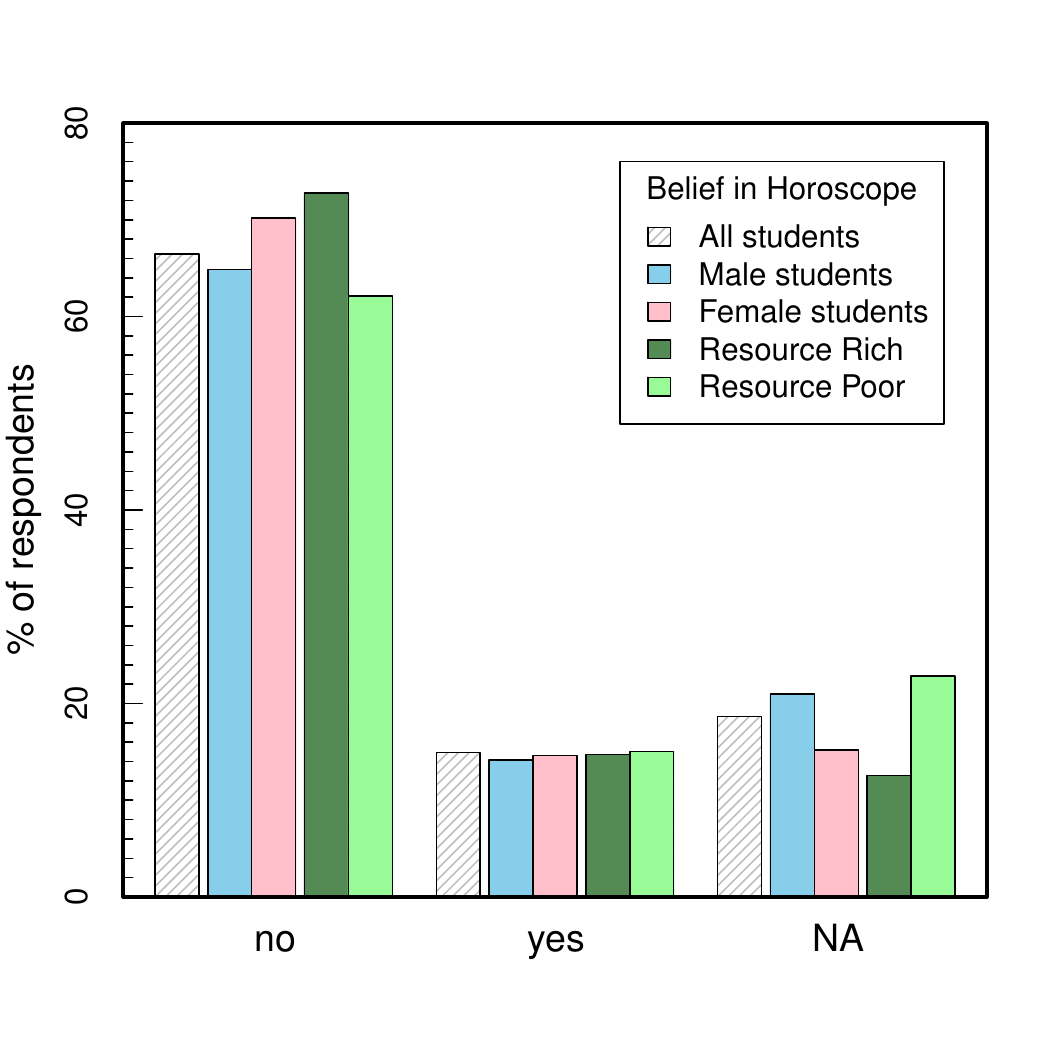}
    \caption{Histograms of percentages of student responses to the hypothetical scenario: "Ramesh cancelled his vacation travel plans after reading his horoscope. Would you do the same?" Non-response is indicated as 'NA'. The different colours represent different demographics, all students (grey), male (blue) and female (pink) students, along with students from resource-rich (dark green) and resource-poor (light green) schools.}
    \label{fig:horo}
\end{figure}

In India,  a large fraction of people believe in astrology. Hence we aimed to explore this aspect among the student population through our survey. To gauge the prevalence of this belief, we adopted an indirect approach, recognizing the potential for response bias in direct questioning. Instead of a straightforward yes/no question, students were presented with a hypothetical scenario: "Ramesh cancelled his vacation travel plans after reading his horoscope. Would you do the same? Explain your answer." This situational question aimed to elicit more genuine responses reflective of students' true beliefs. Analysis of the responses revealed that $66\%$ of students expressed their reluctance to cancel travel plans based on their horoscope, indicating a notable skepticism towards astrology among this cohort (Figure~\ref{fig:horo}). Conversely, $15\%$ of students indicated they would indeed cancel plans under similar circumstances, while  $19\%$ refrained from responding to the question. This distribution suggests a relatively positive outlook, with approximately two-thirds of students expressing a healthy skepticism about astrology or horoscopes.

Upon scrutinizing the rationale behind their responses, discernible trends become apparent. Among students who were unwilling to alter their travel plans, the predominant reasons cited were disbelief in astrology and its characterization as superstition, fallacious, unscientific, or irrational. Conversely, students inclined to alter their plans tended to justify their decision by asserting that horoscope is accurate, it has a mysterious influence on daily life, and that a large population around them believes in astrology.

Further examination of the data revealed minimal, statistically insignificant differences between genders in their responses. However, disparities along socioeconomic lines were slightly more pronounced, with $73\%$ of students from resource-rich schools expressing skepticism towards astrology, compared to $62\%$ of students from resource-poor schools. Despite these variations, the overarching trend suggests prevalent skepticism towards astrology among the student population.

%\begin{figure}
%    \centering
%    \includegraphics[width=0.45\textwidth]{figures/horo_pie_data_v2.pdf}
%    \caption{Students' belief in astrology}
%    \label{fig:horo}
%\end{figure}

\subsection{Access to Astronomy Resources}
% Have you ever looked at the night sky? Approximately how many stars could you see?

In our survey, we also explore if the students have access to astronomical resources, which include natural resources such as a clear night sky, and modern resources, e.g., telescopes, planetariums, and anything else they might use to gain more astronomy-related knowledge.

\subsubsection{Night sky}
\begin{figure}
    \centering
    \includegraphics[width=0.5\textwidth]{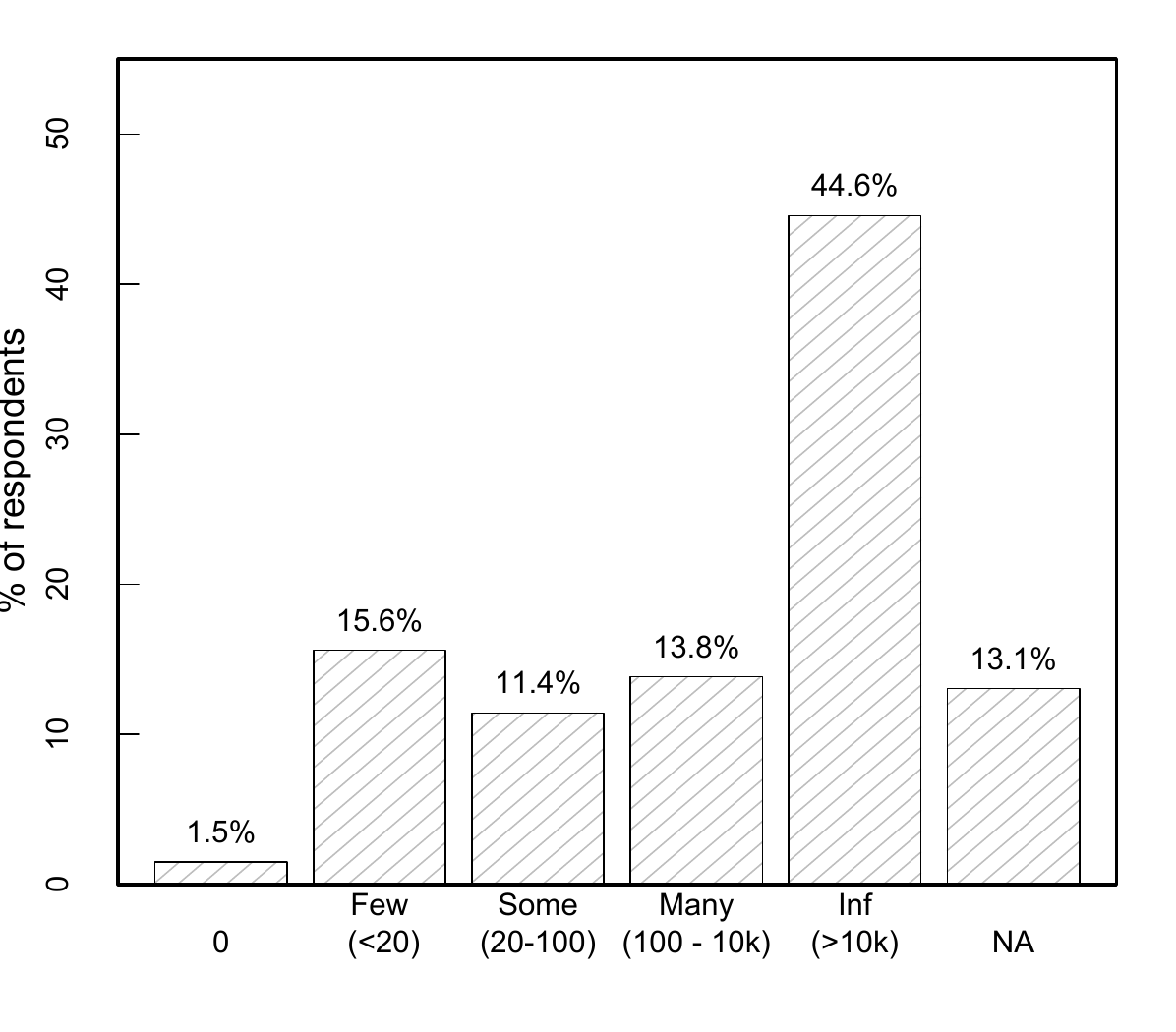}
    \caption{Number of stars seen in the sky as reported by the students.}
    \label{fig:nstars}
\end{figure}

Students were prompted to estimate the approximate number of stars they had observed in the night sky. The textual responses were subsequently categorized into groups: zero, few (defined as < 20), some (20 - 100), many (100 - 10,000), and infinite (> 10,000),  with the corresponding distribution illustrated in Figure~\ref{fig:nstars}. We find that the most popular response ($45\%$ of students) indicate unrealistically large number of stars observed. This is followed by options such as a few stars, many stars, and some stars, all ranging within a response rate of $11 - 16\%$. We note that these figures are quite similar for students from resource-rich and resource-poor schools. In actuality, in many places in India, especially in big cities like Mumbai / Pune, we can hardly see even a few tens of stars, due to high levels of light and air pollution. Consequently, responses indicating the sighting of infinite or tens of thousands of stars are likely to reflect textbook knowledge regarding countless stars in space rather than first-hand experiences. It is possible that some of the students may have misunderstood the question and answered ``How many stars are (supposed to be) in the sky?''. However, this again shows an expectation on student's part to be asked about their textbook knowledge and not about their lived experience. We also note if the students have mentioned anything about the effect of pollution on the stargazing experience, and find that only $7\%$ of respondents mention this, although this awareness is more prevalent among students from resource-rich schools ($12\%$) compared to their counterparts ($4\%$). No discernible differences were observed among genders in this regard.

\subsubsection{Telescopes and Planetariums}

\begin{figure*}
\centering
    \includegraphics[width=0.45\textwidth, trim=10 50 10 100]{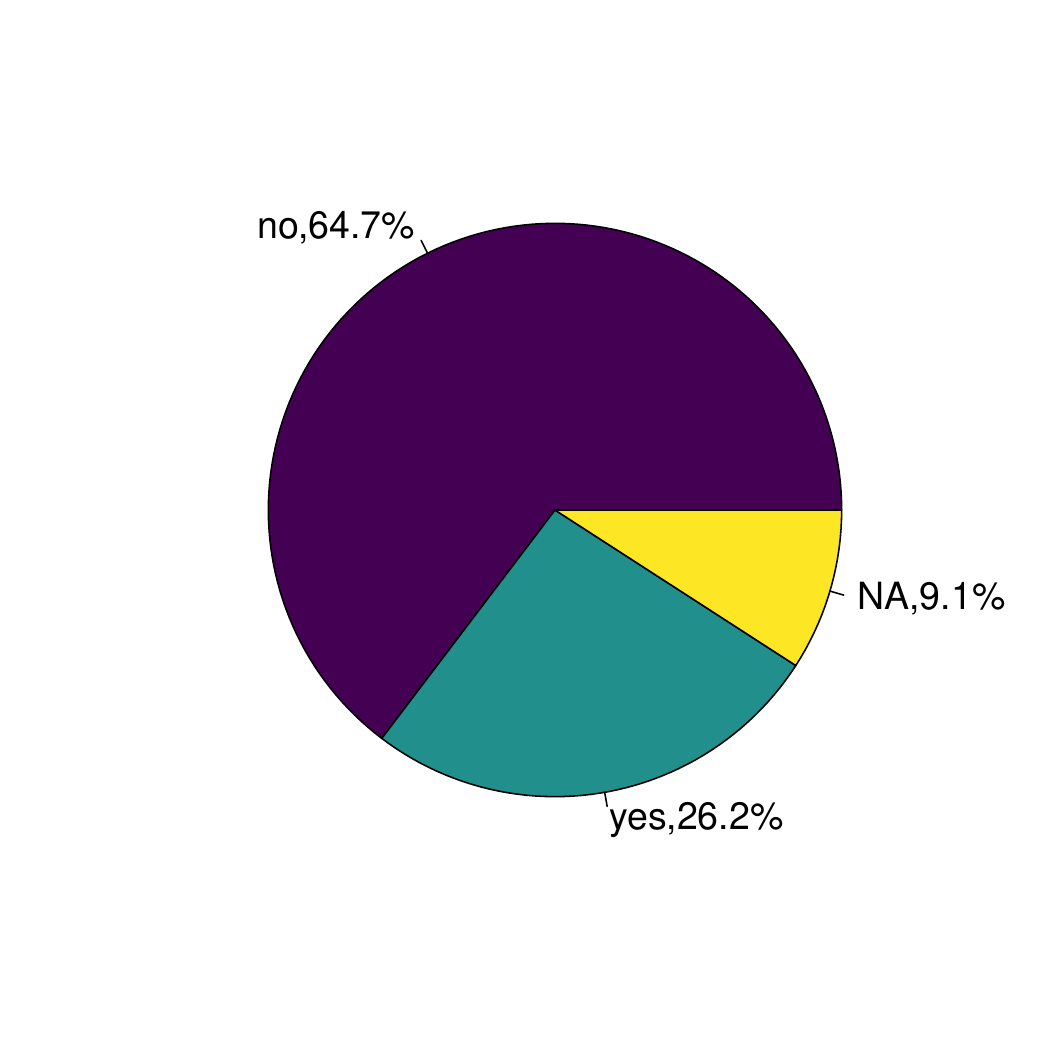}
    \includegraphics[width=0.45\textwidth, trim=10 50 10 100]{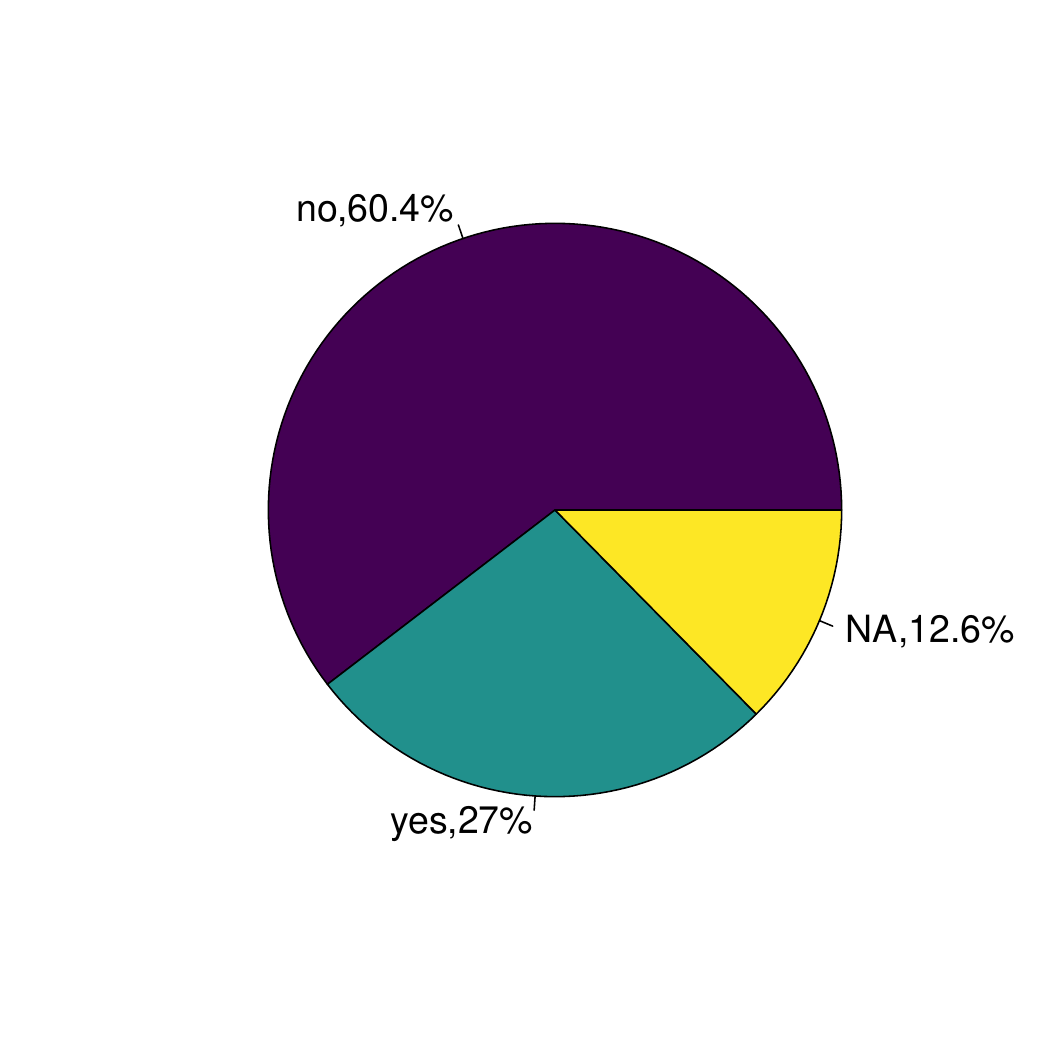}
    \caption{Pie chart of student responses for access to telescopes (left) and planetariums (right). Non-response is indicated as 'NA'.}
    \label{fig:pie_tel_pla}
\end{figure*}

\begin{figure}[h]
    \centering
    \includegraphics[width=0.45\textwidth]{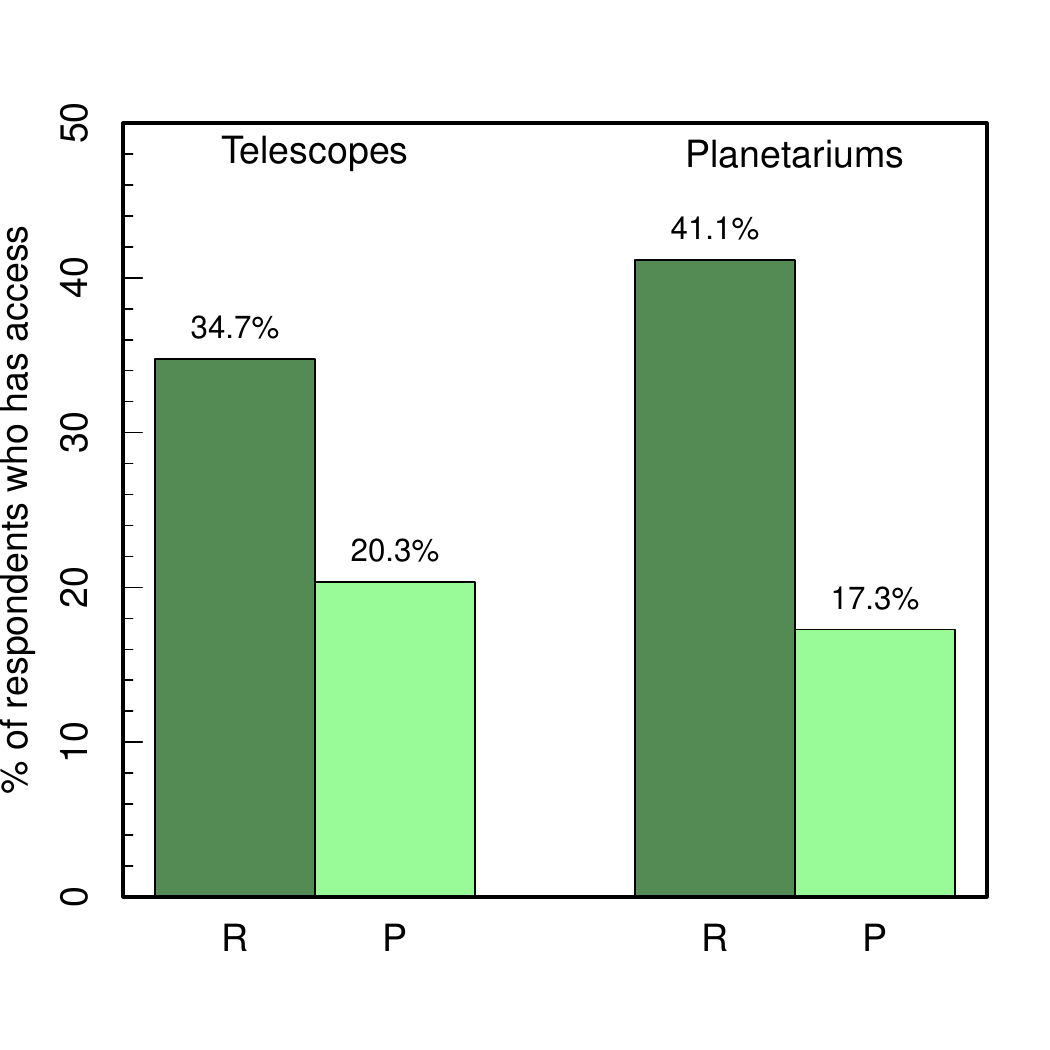}
    \caption{Comparison of percentage of students who have access to Telescopes (left) and Planetariums (right). Here R and P are abbreviations for Resource-rich and Resource-poor school students respectively.}
    \label{fig:rp_tel_pla}
\end{figure}
In the survey, we inquired whether students have ever engaged in sky observation through a telescope. The data reveals that only 26$\%$ of survey participants have ever done so (Figure~\ref{fig:pie_tel_pla}). A majority of respondents (65$\%$) responded negatively, while 9$\%$ did not answer this query. Further analysis exposes significant disparities along socioeconomic divides, with 35$\%$ of students from resource-rich schools reporting telescope access compared to a lower rate of 20$\%$ among their resource-poor counterparts (Figure~\ref{fig:rp_tel_pla}).

Similarly, when queried about visits to a planetarium, the responses exhibit a similar pattern, with 27$\%$ of all respondents having ever visited one (Figure~\ref{fig:pie_tel_pla}). Furthermore, among students from resource-rich schools, this figure rises to 41$\%$, contrasting starkly with the much lower rate of 17$\%$ observed among resource-poor school students (Figure~\ref{fig:rp_tel_pla}). Additionally, anecdotal observations during informal discussions with students revealed that many students are not familiar with the concept of planetariums at all. These findings collectively underscore the widespread scarcity of access to astronomical facilities across the country, with pronounced discrepancies evident across socioeconomic lines.

%\begin{figure*}
%    \centering
%    \includegraphics[width=0.33\textwidth]{figures/tel_pla/pla_pie_data_v2.pdf}
%    \includegraphics[width=0.33\textwidth]{figures/tel_pla/pla_rich_pie_data_v2.pdf}
%    \includegraphics[width=0.33\textwidth]{figures/tel_pla/pla_poor_pie_data_v2.pdf} 
%    \includegraphics[width=0.33\textwidth]{figures/tel_pla/tel_pie_data_v2.pdf}
%    \includegraphics[width=0.33\textwidth]{figures/tel_pla/tel_rich_pie_data_v2.pdf}
%    \includegraphics[width=0.33\textwidth]{figures/tel_pla/tel_poor_pie_data_v2.pdf}
%    \caption{Access to Planetarium (top) and Telescopes (bottom) for all students (left), students from resource-rich schools and resource-poor schools.}
%    \label{fig:tel_pla}
%\end{figure*}

\subsubsection{Other resources}
\begin{figure}
    \centering
    \includegraphics[width=0.45\textwidth]{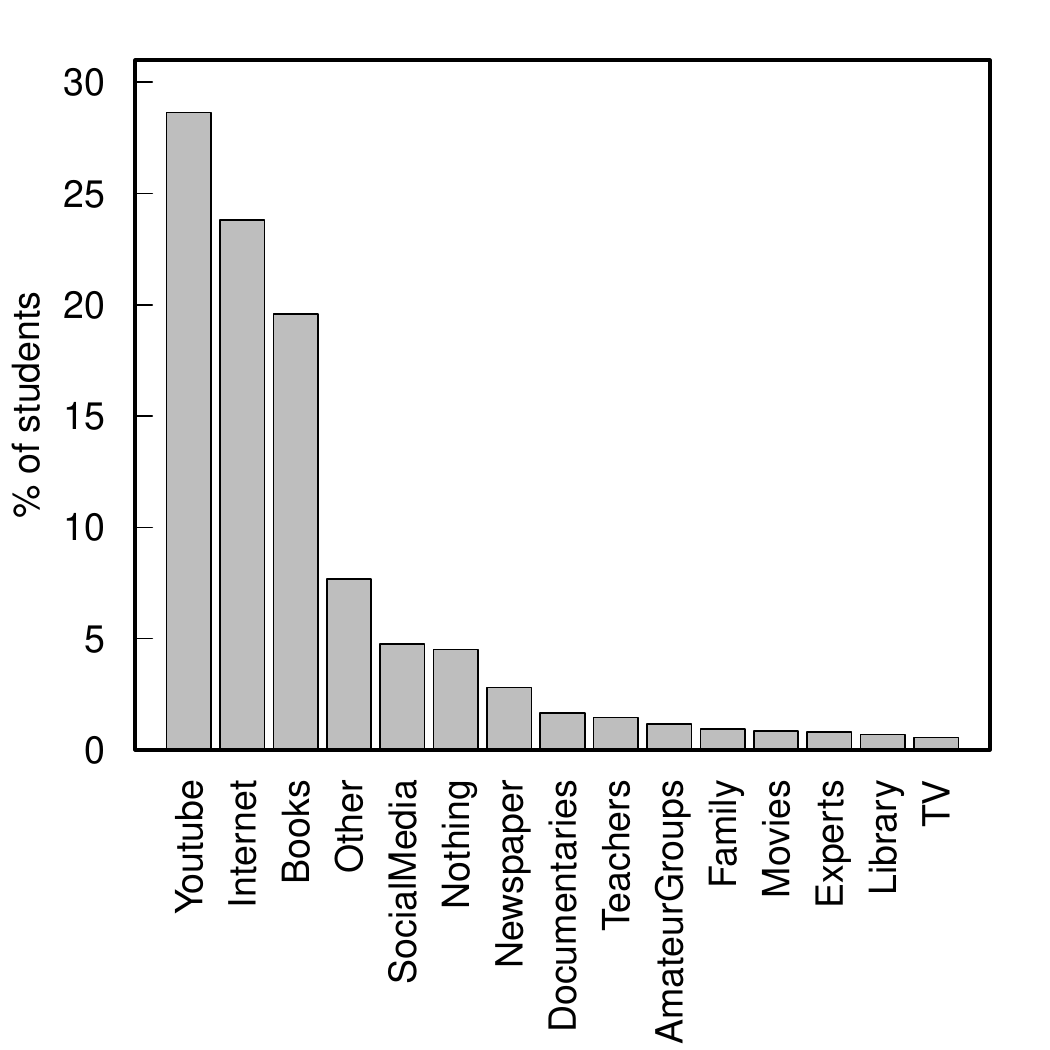}
    \caption{Percentages of students who prefer different resources for gaining further astronomy knowledge.}
    \label{fig:other_source}
\end{figure}

In the survey, we also inquired about supplementary resources utilized by students to augment their understanding of astronomy beyond the confines of their textbooks. We categorized the various textual responses, and the distribution of various resources used by students is depicted in Figure~\ref{fig:other_source}. Notably, the predominant resource cited by approximately 29$\%$ of students is YouTube (and other similar video portals). Following closely are the Internet (this keyword covers all websites with textual material as well as web searches) and books, mentioned by approximately 24$\%$ and 20$\%$ of students, respectively. Conversely, resources such as guidance from teachers, experts, or exposure through mass media channels (television, documentaries) are cited by less than 5$\%$ of students. This analysis unequivocally underscores the preference and/or prevalence among students for online resources, particularly YouTube, as the primary medium for acquiring knowledge about astronomy. Nonetheless, it is imperative to acknowledge that while many YouTube channels offer valuable content, many others propagate misinformation and superstitions. Regrettably, the present school curricula do not address the crucial aspect of teaching students how to discern between credible and unreliable online resources. Therefore, there is an urgent need to equip students with such skills in schools.

\subsection{Interest towards astronomy}
%\textbf{Do you like learning about astronomy in general?}
The survey queried students if they liked learning about astronomy in general and we found that the vast majority of students, specifically $85.9\%$, like the subject of astronomy. Further $4.1\%$ students are ambivalent towards the subject, while only $10.1\%$ students expressed a negative sentiment. Further analysis shows that such high levels of interest are common in all subgroups of students, e.g. among males, females, and students from resource-poor and resource-rich schools.

%\textbf{What other astronomy topics would you like to learn about?}

\subsubsection{Astronomy topics of interest}
\begin{figure}[h]
    \centering
\includegraphics[width=0.45\textwidth]{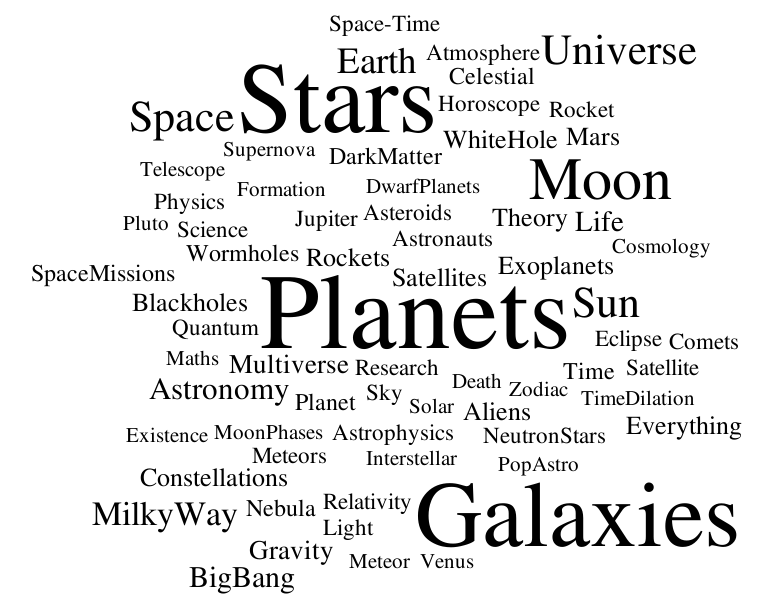}
    \caption{Word cloud depicting topics students expressed interest in learning more about, as indicated in the survey responses.}
    \label{fig:topics_learn}
\end{figure}

Students were queried regarding their interest in additional astronomy-related topics they would like to explore. The responses yielded a diverse array of textual data, visually represented in Figure~\ref{fig:topics_learn} in the form of a word cloud. Analysis of the data reveals that prevalent areas of interest among students include planets, stars, and galaxies. Moreover, popular science topics such as the Big Bang, the Universe, space, the Multiverse, and Black holes emerged as recurrent themes. We note that most syllabi across the country lack substantial content on galaxies, stellar astronomy, and cosmology. Drawing from these findings, it is evident that students harbor a keen interest in delving deeper into these topics, which are integral components of school curricula in many other countries \citep{Salimpour2021}. Consequently, it is recommended that consideration be given to incorporating these topics into the curriculum in the future, at least as an astronomy elective course.

%\textbf{Would you like to learn more about astronomy in higher classes or college?}

\subsubsection{Astronomy in higher studies}
Students were also queried regarding their interest in pursuing further studies in astronomy at higher classes or in college. Analysis of the responses revealed that a substantial majority, comprising 70$\%$ of students, expressed a positive interest in such endeavors while 6$\%$ expressed ambivalence. Conversely, a minority of respondents (18$\%$) indicated a negative disposition towards pursuing further studies in astronomy. We note that a small proportion of students (7$\%$) refrained from responding to this query. Further analysis shows us that while $72\%$ female students are interested in further astronomy learning, only $66\%$ male students are interested in doing so. On the other hand, $72\%$ of resource-poor school students respond positively compared to $66\%$ students from resource-rich schools.

\subsubsection{Careers in Astronomy}
Upon querying students regarding their aspirations to pursue a career as an astronomer, our findings reveal that 40$\%$ of students responded affirmatively, with an additional 8$\%$ expressing indecision. Conversely, 45$\%$ of students indicated a negative response, while 8$\%$ abstained from answering the question. Upon analyzing this data along socioeconomic lines, it emerges that among students from resource-poor schools, 46$\%$ expressed a desire to become astronomers, compared to a lower rate of 30$\%$ among their resource-rich counterparts. This discrepancy may initially appear perplexing, particularly in light of our previous discussion regarding access to astronomical resources, where we observed that resource-rich schools typically enjoy greater access to telescopes, planetariums, etc., factors traditionally associated with a heightened inclination towards a career in astronomy. However, anecdotal evidence gleaned from informal conversations with students provides insight into this phenomenon. It appears that students from resource-rich schools often possess predetermined career aspirations by this age (respondents age is typically 14 - 15), gravitating towards professions such as engineering, administrative roles, or medicine, with a notable lack of interest in pursuing research-based careers. Conversely, students from resource-poor schools may not have their career paths fully defined at this age, affording them more flexibility in their aspirations, thus fostering a greater inclination towards astronomy as a potential career path.

Finally, we inquire about students' understanding of the pathway to becoming an astronomer. Rather than posing the question directly, we present a scenario wherein an individual named Asha aspires to pursue a career in astronomy and ask what course of study she would need to undertake after completing school and college. However, our findings indicate a pervasive lack of clarity among students in this regard. The majority of respondents simply suggested studying science, physics, or astronomy without providing specific details. Informal discussions with students and their teachers corroborated these findings, revealing a widespread lack of awareness regarding the educational and career trajectory required to become an astronomer, even among those expressing an interest in the field. These results emphasize the need for improved career guidance in schools and better-informed educators to support students in understanding diverse career paths, including astronomy. It is plausible that the lack of awareness and guidance surrounding science-related career paths contributes to the overall disinterest observed among students toward science and research-based careers.

%\begin{figure*}
%    \centering
%    \includegraphics[width=0.33\textwidth]{figures/astronomer/astronomer_pie_data_v2.pdf}
%    \includegraphics[width=0.33\textwidth]{figures/astronomer/astronomer_rich_pie_data_v2.pdf}
%    \includegraphics[width=0.33\textwidth]{figures/astronomer/astronomer_poor_pie_data_v2.pdf}
%    \caption{Pie chart showing responses to the question if the students want to be an astronomer in the future. Responses from all students, students from resource-rich schools, and resource-poor schools are shown in left, middle, and right respectively.}
%    \label{fig:enter-label}
%\end{figure*}

%\subsection{Effect of gender differences in the data}

%\subsection{Effect of socioeconomic factors in the data}

\section{Discussions and Inferences}
\label{sec:discussion}

This survey is certainly first of its kind among the emerging countries in Asia. In fact, this is a first survey anywhere covering both kinds of classrooms: the ones comparable to first world countries and the ones comparable to underdeveloped regions. Further, this is one of the few science education studies involving sample size of more than 2000 respondents. In these respects, this survey brings out several new trends to light and creates opportunities for further explorations.

Our analysis unveils some interesting trends within the surveyed population. While male students exhibit greater proficiency in knowledge-based inquiries concerning celestial sizes and distances, female students perform slightly better in answering questions about lunar phases, which encompass both observational and conceptual components. Notably, female respondents are much more aware of the connection of astronomy with their cultural heritage. They also express a stronger inclination toward further astronomical studies and pursuing careers in the field.

Likewise, students hailing from resource-poor educational setups generally perform less favorably across all astronomy knowledge domains. Their access to resources is severely limited, compared to their counterparts from resource-rich schools. Despite this, they are also more aware of the cultural significance of astronomy and express considerably more interest in expanding their knowledge of the subject and pursuing careers in astronomy. 

Our analysis highlights interesting parallels between the responses of female respondents and students from resource-poor backgrounds. It must be noted that female students are evenly distributed across both resource-rich and resource-poor schools. This indicates that societal norms place female students in a similar disadvantaged position as the learners in resource-poor environments. However, lower societal expectations also allow these groups to do non-formal explorations of the world around them. In future, we aim to explore this aspect in more detail to understand these sociological dynamics better.

As mentioned in section \ref{sec:methods}, the data was collected in a large number of states across India. These states individually also are at different rungs of human development index and often follow different textbooks covering similar content. In future, it would be worthwhile to explore this axis of diversity in our sample as well.

In a way, the diversity along multiple axes also poses a challenge and may be construed as a limitation of our study. One may take the view that the sample or the identified subsamples (along gender or socioeconomic axes) are not homogeneous enough to analyse them as a single group. However, we argue that the trends observed in our data are so pervasive that they must be present in all subsamples. In that sense, the student sample is a good representation of the overall student population across the country. In future, we hope to include all the remaining states and compare the statewise results to get an understanding of relative teaching-learning standards.

%different states?
%
%\subsection{Conclusions}
%\label{sec:conclusions}

Returning to the lines of our research inquiries, our investigation unveils major deficiencies in students' comprehension of basic astronomical concepts despite exposure to instructional materials. Nonetheless, students harbor a profound fascination with astronomy, with a sizable proportion expressing interest in learning further astronomy in higher studies and even pursuing careers in the field. Less than half of the students seem to recognize the close connection between their cultural roots and astronomy, although, on the positive side, a majority reject astrological beliefs. Gender and socioeconomic disparities are apparent, with the latter exhibiting greater significance across various metrics.

In summation, our findings advocate for capitalizing on students' innate curiosity towards astronomy to include more astronomy content in a more meaningful way. These insights can serve as a foundation for the development of tailored interventions, including targeted teacher training programs, and refined curriculum frameworks, aimed at bolstering astronomy education in secondary schools. These programs can be implemented through formal, non-formal and hands-on learning under the provisions of National Education Policy in India.

The results of this survey provide data-driven insights into student minds that can help improve educational curricula, textbooks and other curricular materials, and teaching practices. Additionally, this study fills a clear gap left by a scarcity of large-scale surveys focusing on developing countries. Thus, our study is not just useful for Indian classrooms but can also offer valuable insights that are applicable to similar educational contexts across many countries in Asia, Africa and South and Central America.

\section{Declarations}

\subsection{List of abbreviations}

AER = Astronomy Education Research, ASSCI = Astronomy and Space Science Concept Inventory, CI =  Concept Inventory, MCQ = Multiple Choice Question, NCERT = National Council of Educational Research and Training, OAE =  Office of Astronomy for Education, IAU =  International Astronomy Union.  

\subsection{Ethical Approval}
The research proposal and questionnaire was presented to the ethics board at the institute of one of the co-authors, namely at the Homi Bhabha Center for Science Education, Mumbai.  After a rigorous evaluation process, we received clearance from the ethics board and the primary data collection commenced only after the approval.

\subsection{Consent for publication}

 The survey is completely anonymous and entirely voluntary. We took explicit informed consent from all participants; students, teachers, and school principals. In case of students, in addition to the consent form in the questionnaire, we also explained the importance of their consent verbally in each classroom before the start of the survey. For school principals / teachers-in-charge, the relevant consent form was signed by both them and the research team. 

\subsection{Competing Interests}

The author(s) declare that they have no competing interests.

\subsection{Funding}

No external grant was used to conduct this survey. The minor printing and travel costs were covered by parent institutes of the collaborators.

\subsection{Author's Contributions}

The baseline survey was conceptualised and created by the first three authors (M. Maji, S. More and A. Sule) with inputs from current and former colleagues at IUCAA and HBCSE (P. Ranadive, S. Shetye, A. Bhandari). The same team was also responsible for the pilot study, its analysis, the second iteration of the survey instrument and also the design of necessary consents, data collection protocols and data encoding protocols. M. Maji performed most of the data analysis (with inputs from S. More and A. Sule) and also wrote the original draft of the manuscript. S. More and A. Sule were involved in structuring the flow of the manuscript, refining the original draft and overall project supervision. All other co-authors were involved in translation of questionnaire, data collection in respective states and coding their share of data as per the prescribed scheme. All co-authors reviewed the manuscript draft and gave detailed suggestions for improvement.

\section{Acknowledgements}

We sincerely thank all the school principals, teachers and students who took part in our survey.

We also acknowledge the valuable contribution of all the other volunteers who helped with our survey. They are Pooja Joshi and  Arjun Singh (both from ARIES, Nainital), Puneeth R. and Snehalata (both from IIA and COSMOS Mysuru), Masoom Jethwa (Vikram A Sarabhai Community Science Centre), Altamash Shaikh, Pooja Tolia, Deep Samanta and Sharmila Budhbhatti (volunteers from the Sky Explorers). We additionally thank Dr. Shamin Padalkar (TISS) for many valuable discussions about this project and astronomy education research in general.

\section{Authors' information}

The first three authors (M. Maji, S. More, and A. Sule) are all trained astronomers who are starting to work in astronomy education and AER at school level. The rest of the team includes trained astronomers, astronomy educators, active school teachers as well graduate students.

\bibliography{main_paper.bib}

\appendixpage

\section{Appendix 1: Baseline Survey Questionnaire}
\label{sec:appendix}

The Baseline Survey Questionnaire is presented in the following pages. It is the English version of the survey with some terms also written in the Hindi / Marathi script. This version was used in English medium schools. Page 2 of the survey was kept intentionally blank (to separate the demographics part of the survey from the responses part, if needed), so that page is not shown here.

\clearpage
\newpage



\section*{Supplementary material}

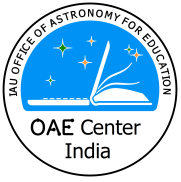

**International Astronomical Union (IAU)**
**Office of Astronomy for Education (OAE) Center - India**

# *Baseline Survey of Astronomy* Education in India*

## Consent Form for Students

**Dear student,**

**This is not an exam.** We are doing research on the state of astronomy education in India. For this, we are collecting information about what students and teachers think about astronomy content in the school syllabus. Please answer the questions on your own. Please let us know if you are uncomfortable answering any of these questions.

We will collect similar responses from students from many schools across India and present overall statistics in a report. At no stage, your name or any other personal details will be revealed. Your participation is voluntary. You agree to be part of this survey by choosing to answer the questions **in the survey (page 3 onwards).** At any stage, you may choose to opt-out of this survey. We will remove and destroy your response sheet if you opt out.

*Astronomy is the study of the objects outside the earth (like planets, stars, galaxies) as well as various motions of earth and astronomers are the scientists studying astronomy.

## Participant details:

School name - _____

Standard and section -_____

Roll No. - _____

Gender - Female / Male / Other


City/ Village - _____

Taluka - _____

District - _____

State - _____

# Questionnaire for students:

1) What is your favourite subject?

2) Do you like learning about astronomy in general? (see first page if you are not familiar with the word astronomy)

3) Among the astronomy topics you have <u>learned in school so far</u>,

   a) which topics interested you the most?

   b) which topics interested you the least?

   c) which topics did you find the hardest?

4) Arrange the objects below from the nearest to farthest (from the earth).

   Sun, Moon, Stars, Neptune

5) Arrange the objects below from the smallest to the largest.

   Jupiter, Moon, Earth, Sun

6) Choose the correct image for the given phases of the moon.

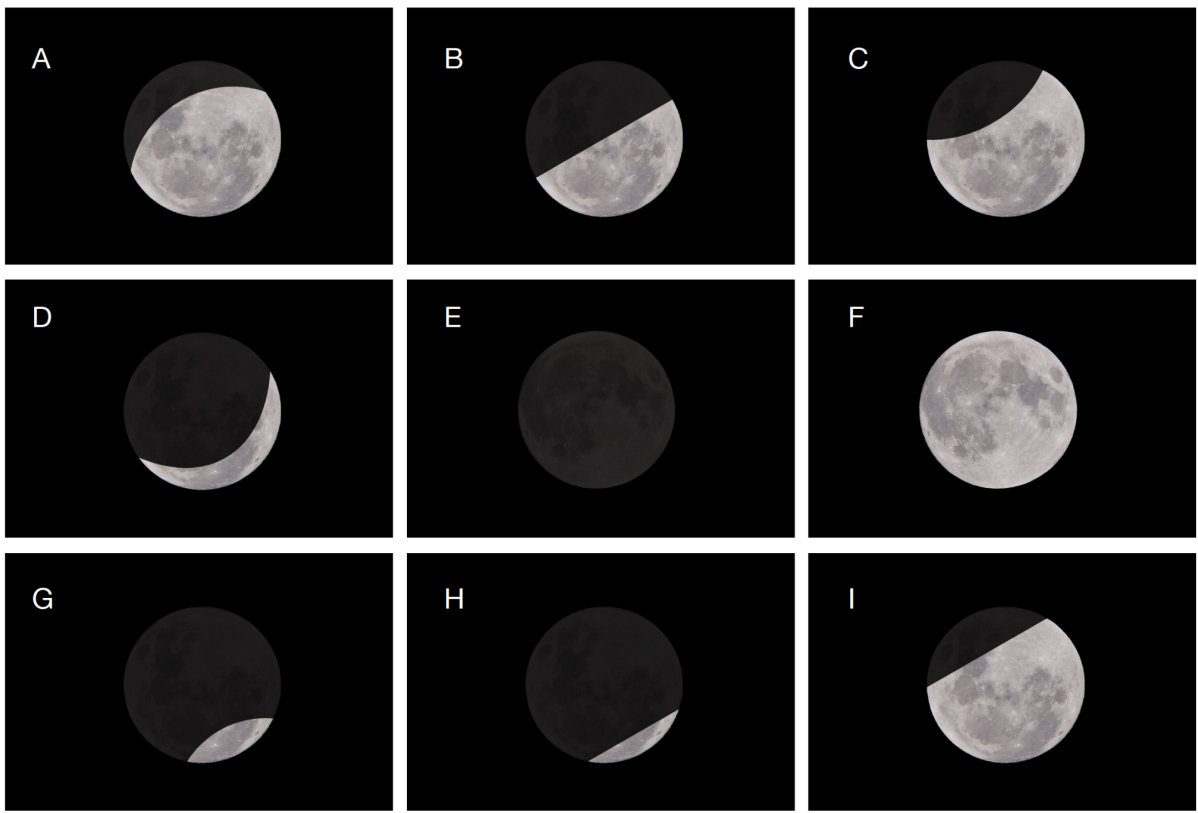

   Moon Phase                                        Image

   New Moon (Amavasya, Day 1)    -
   2nd Day                       -
   8th Day                       -
   11th Day                      -
   Full Moon (Purnima)           -

7) Name some festivals that always fall on

    a) full moon day (Poornima / पौर्णिमा ) -

    b) new moon day (Amavasya / अमावस्या ) -

8) Ramesh cancelled his vacation travel plans after reading his horoscope (राशीफल / राशीभविष्य). Would you do the same? Explain your answer.

9) What other astronomy topics would you like to learn about?

10) Apart from your textbook, what other sources have you used to know more about astronomy? Write names of books, videos, etc.

11) Have you ever looked at the night sky? Approximately how many stars did you see?

12) Have you ever looked at the sky through the telescope? What did you see?

13) Have you ever visited a planetarium / mobile planetarium? If yes, give some details about your experience.

14) Would you like to learn more about astronomy in higher classes or in college?

15) Would you like to be an astronomer?

16) Asha wants to become an astronomer. To do that, what course of study would she have to choose,

After 10th class?

After 12th class?

After college?

Anything else?

—------ End of Survey —-----



\end{document}